\documentclass[acmsmall,natbib=false]{acmart}
\usepackage{enumitem} %
\usepackage[ruled,linesnumbered]{algorithm2e}
\usepackage{multirow}
\AtBeginDocument{%
  }

\acmJournal{TOIS}

\acmVolume{1}
\acmNumber{1}
\acmArticle{1}
\acmMonth{8}

\SetKwInOut{Require}{Require}
\SetKwInOut{Ensure}{Output}
\RequirePackage[
  datamodel=acmdatamodel,
  style=acmnumeric,
  ]{biblatex}

\begin{document}

\title{Joint Workflow and Prompt Optimization for User Behavior Simulation}
% Task-Agnostic
\author{Nipun B Nair}
\authornote{primary author}
\email{nipun.nair@monash.edu}
\orcid{0009-0000-7342-7732}
\author{Tongtong Wu}
\email{tongtong.wu@monash.edu}
\orcid{0000-0002-5752-9779}
\affiliation{%
  \institution{Monash University}
  \city{Melbourne}
  \state{Victoria}
  \country{Australia}
}
\author{Hongzhi Yin}
\email{db.hongzhi@gmail.com}
\orcid{0000-0003-1395-261X}
\affiliation{%
  \institution{The University of Queensland}
  \city{Brisbane}
  \state{Queensland}
  \country{Australia}
}
\author{Hui Li}
\email{hui@xmu.edu.cn}
\orcid{0000-0001-9139-3855}
\affiliation{%
  \institution{Xiamen University}
  \department{Department of Computer Science and Technology}
  \city{Xiamen}
  \country{China}
}
\author{Weiqing Wang}
\email{teresa.wang@monash.edu}
\orcid{0000-0002-9578-819X}
\affiliation{%
  \institution{Monash University}
  \city{Melbourne}
  \state{Victoria}
  \country{Australia}
}

\renewcommand{\shortauthors}{Nipun et al.}

%% article. task-agnostic
\begin{abstract}
User behavior simulation is the computational modeling of user interactions within information systems through the use of simulated agents in place of live users. It supports system testing and evaluation, decision-making and forecasting, and user experience design. Existing simulators rely on hand-crafted rules or domain expertise that transfers poorly across tasks.  \textbf{SWORD} (Simulation-driven Workflow and Prompt Optimization with Role-based Design) is introduced as a framework that jointly optimizes multi-agent workflow topology and natural-language prompts. It is guided solely by a scalar task metric, without domain initialization or task-specific engineering. The experimental results demonstrate that SWORD achieves statistically significant gains over prompt-only, workflow-only, and staged-optimization baselines under a controlled, identical-backbone comparison. Against the strongest published domain-specific baseline, SWORD further improves accuracy while using a smaller backbone model, substantially less training data, and a very reasonable API cost (\$4--\$6 for each dataset). 
%Because this comparison spans different backbone models, we report it as a cost-efficiency result rather than a controlled accuracy comparison. 
Beyond predictive performance, SWORD autonomously discovers domain-relevant signals, review-sentiment mapping rules and epidemiological decay priors, purely from scalar error feedback, establishing textual gradients as a mechanism for unsupervised feature-importance discovery in user behavior modeling.
\end{abstract}

% \received{1}
% \received[revised]{1}
% \received[accepted]{1}

\maketitle
\section{Introduction}
\label{sec:intro}

\begin{figure}[htbp]
  \centering
  \includegraphics[width=\linewidth]{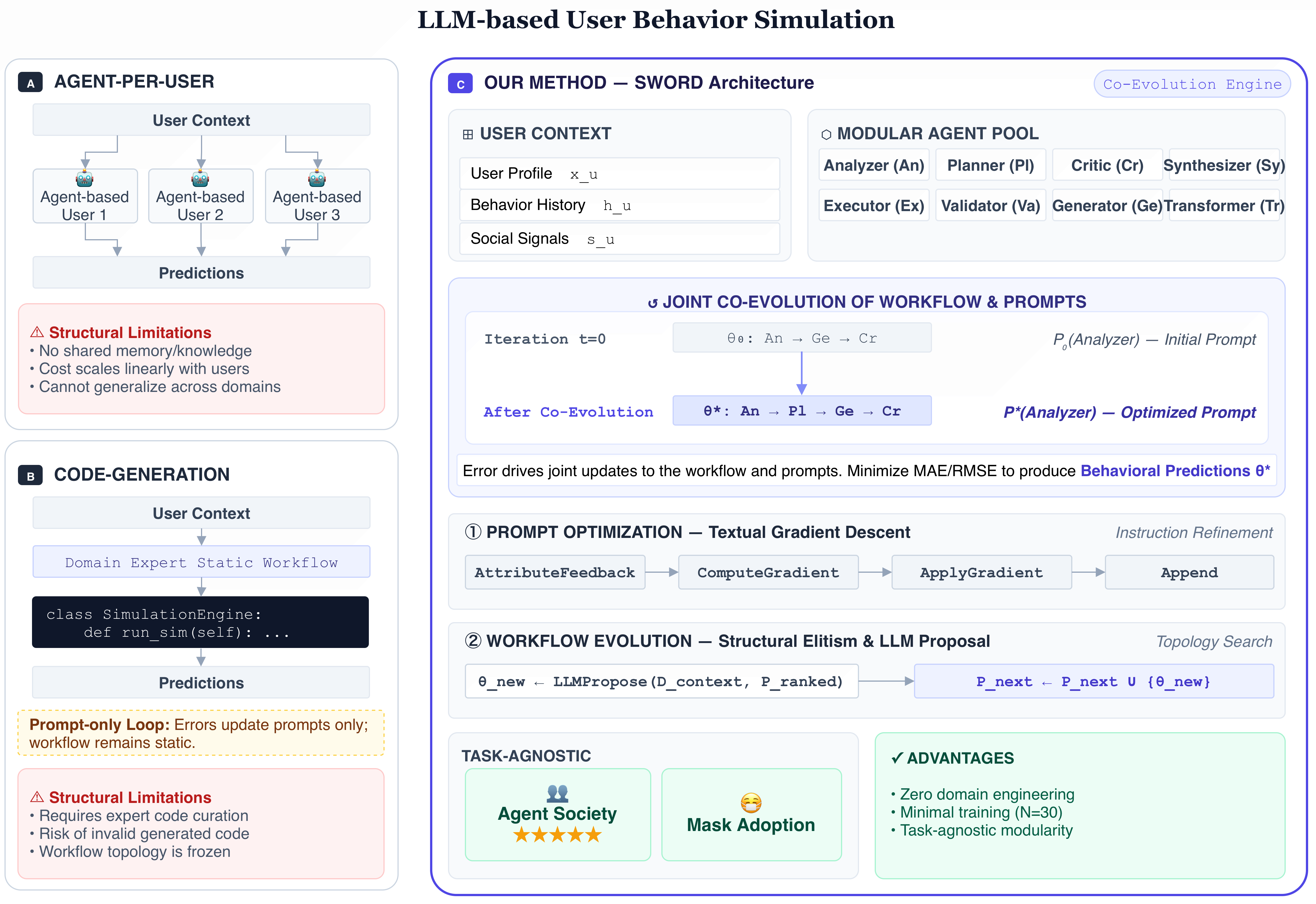}
  \Description{A three-panel diagram comparing paradigms of user behavior simulation. Panel A shows the Agent-per-User paradigm with isolated LLM agents. Panel B shows the Code-Generation paradigm where an LLM generates a shared Python execution function. Panel C shows the SWORD paradigm featuring a shared workflow DAG (Analyzer, Planner, Generator, Critic, Synthesizer) optimized via an interleaved co-evolution loop of workflow evolution and prompt optimization.}
  \caption{Comparison of LLM based user simulation paradigms and the SWORD framework. 
  \textbf{(A) Agent-per-User}: instantiates isolated LLM agents per user, preserving personalization but sharing no knowledge. 
  \textbf{(B) Code-Generation}: achieves personalization via generating executable Python simulation code for each user through shared agent workflow, while shifting the burden of workflow design and simulation logic validation to domain experts, limiting task transferability. 
  \textbf{(C) SWORD (Ours)}: is a further step forward compared with Code-Generation to eliminate the dependence on domain experts for different user simulation tasks via automated workflow and prompt optimization and task agnostic modular agent pool design. }
  %Learns a single shared multi-agent workflow ($\theta^*$) that operates as a universal function mapping user context to shared behavioral predictions. Rather than optimization in isolation, SWORD is task agnostic and does not require expensive domain expertise for design.}
  \label{fig:paradigms_and_framework}
\end{figure}
% \begin{figure}[t]
%   \centering
%   \includegraphics[width=\linewidth]{building_blocks.png}
%   \Description{A two-panel conceptual diagram. Panel 1 explains the core concepts: workflows are visualized as structural processes or blueprints determining the interaction structure between agents, while prompts are visualized as instruction cards providing explicit contextual guidance to an LLM. Panel 2 shows how these concepts apply to user simulation: a shared optimized multi-agent workflow maps user context through a sequence of roles (Analyzer, Planner, Generator, Synthesizer) to a final behavior prediction, while a detailed breakdown compares a basic generic prompt with a highly optimized, metric-anchored SWORD prompt.}
%   \caption{Conceptual breakdown of simulation building blocks. 
%   \textbf{(1) Concepts}: The workflow serves as the structural process (blueprint) mapping information flow and agent communication topology, while the prompt acts as the individual instruction set (contextual guidance) dictates how each agent behaves within a node. 
%   \textbf{(2) Simulation Application}: During behavioral simulation, the shared workflow framework dynamically routes user history, profiles, and social signals through task-agnostic roles. Simultaneously, SWORD refines the generic prompts into metric-anchored, highly specialized instructions to accurately capture complex user actions and preferences.}
%   \label{fig:building_blocks}
% \end{figure}
User behavior simulation holds significant value in information systems in three key ways. First, it enables informed decision-making and forecasting by predicting how users will interact with system features before actual deployment\cite{cai2025agentic,park2025llm}. Second, it supports system testing and optimization by examining user behavior across diverse scenarios and contexts\cite{ma2025pub}. Third, it improves user experience design by revealing how different user segments interact with systems, thereby enabling researchers to create more intuitive and responsive interfaces\cite{he2025simulating}. 

Classical approaches to user behavior simulation use probabilistic frameworks, including click models, cascade models, and position-bias models, which are fitted to logged data~\cite{balog2024usersim,dupret2008browsing,maxwell2016simiir}.
Rule-based agent models extend this by encoding explicit decision logic,
interaction rates, and social influence parameters specified by domain
experts~\cite{dekai2020universal,mortveit2022bessie}.
Both families are effective within their original domain, but they transfer
poorly: constructing or adapting them requires expert knowledge that is
rarely available across the full range of tasks that IR and
recommender-system evaluation demands~\cite{balog2024usersim}.

Large language models (LLMs) offer a fundamentally different path.
LLMs encode rich representations of human reasoning, preferences, and
social norms, enabling them to serve as the cognitive backbone of
simulated users without explicit rule
specification~\cite{mact2025cpss,gao2024llmabm,wang2025ubsim}.
Existing LLM-based user simulation approaches can be grouped into two paradigms, each with a
structural limitation (Fig.~\ref{fig:paradigms_and_framework}).
\textit{Agent-per-user} methods~\cite{park2023genagents} instantiate
one LLM agent per simulated user, preserving personalization but sharing
no knowledge across agents: users with sparse interaction histories produce
unreliable predictions, and the system must be rebuilt for each new
simulation domain. Moreover, the number of agents the system needs to
maintain scales linearly with the number of users.
\textit{Code-generation} methods~\cite{hua2026socia} address the scaling
problem by using several agents that communicate through a predefined
workflow to generate Python simulation code: given user profile and history
information, this code predicts behavior as a single shared function.
This removes per-user specialization but shifts the burden onto workflow
design (the workflow that defines how agents communicate) and code
validation which usually depend on domain experts. The generated code requires iterative review and approval from a
domain expert, making transferability contingent on the availability of
domain experts, who are both scarce and costly.
We introduce \textbf{SWORD} (\textbf{S}imulation-driven \textbf{W}orkflow
and prompt \textbf{O}ptimization with \textbf{R}ole-based \textbf{D}esign),
a framework for user simulation that requires no
domain-expert input. SWORD achieves this via joint workflow--prompt
optimization, in which
the structure of the agent workflow (which agents participate and
how they communicate) and the prompts (the instructions given to each agent)
are co-optimized end-to-end in a single closed loop. SWORD takes user
context (profile, behavioral history, social signals) as input and produces
behavioral predictions as output, guided solely by a scalar task metric
(MAE or RMSE).
Each workflow is a directed acyclic graph (DAG) over eight task-agnostic
agent roles: \textsc{Analyzer}, \textsc{Planner}, \textsc{Generator},
\textsc{Critic}, \textsc{Synthesizer}, \textsc{Executor},
\textsc{Transformer}, and \textsc{Validator}.
These roles span the full cognitive space of behavioral prediction, from
data parsing and planning, through generation and critique, to synthesis and
output validation, and they transfer across simulation tasks without
modification.
Adapting SWORD to a new task requires only 30 labeled training examples and a scalar evaluation error metric, with no simulation code, domain initialization, or task-specific engineering.

\noindent\textbf{Contributions.}
\begin{enumerate}[leftmargin=1.8em,itemsep=2pt]

\item \textbf{SWORD: a joint optimization framework for LLM-based user simulation requiring no domain-expert input.}
We introduce a scalable framework that interleaves DAG workflow evolution and prompt optimization within a single optimization loop, with an LLM-guided proposer generating targeted structural mutations informed by the full optimization history. A fixed vocabulary of eight roles transfers unmodified across the two user simulation tasks we evaluate, requiring no domain-expert input or per-task role engineering.

\item \textbf{Workflow--prompt non-stationarity: empirical characterization and efficient search space reduction.} We identify workflow–prompt non-stationarity as an important challenge in multi-agent optimization: the quality of a workflow depends on the prompts it is paired with, rendering staged optimization ineffective. We address this with a practical framework for jointly optimizing agent workflow and prompts, making end-to-end search over both structural and linguistic design spaces computationally feasible. This search space reduction is achieved through mechanisms including elitist selection, a history-conditioned workflow proposer, and worst-case gradient trace selection, as detailed in Section~\ref{sec:search:reduction}.

\item \textbf{Unsupervised user preference discovery via textual gradients.}
SWORD autonomously identifies review text sentiment as the main signal for predicting user ratings. It learns precise rules that map sentiment to rating scores using only MAE feedback. This process establishes textual gradients as a mechanism for unsupervised feature-importance learning in user behavior modeling.
\item \textbf{Improved efficiency through joint optimization.}
SWORD outperforms prior approaches while using a smaller backbone (GPT-5.4-mini) with substantially less training data and computational cost. This highlights that effective joint optimization of workflow structure and prompts can compensate for reduced model capacity and data requirements.
\end{enumerate}

The remainder of the paper is organized as follows.
Section~\ref{sec:related} reviews related work.
Section~\ref{sec:problem} formalises the problem.
Section~\ref{sec:method} presents the SWORD framework.
Section~\ref{sec:experiments} describes the experimental setup.
Section~\ref{sec:results} presents results and
analysis.
Section~\ref{sec:discussion} discusses findings.
Section~\ref{sec:conclusion} concludes the paper.

\section{Related Work}
\label{sec:related}

Our work sits at the convergence of two research streams: user simulation
for information systems, and prompt
and workflow optimization for multi-agent systems. We review each in turn
and identify the shared limitations that SWORD addresses.

\subsection{User Simulation for Information Systems}
\label{sec:related:usersim}

User simulation is a long-established task in information systems, including information retrieval and recommender systems~\cite{balog2024usersim}.
Classical simulators model user–system interactions through probabilistic
frameworks: click models (including the position-bias model, cascade model,
and user-browsing model~\cite{dupret2008browsing}) capture how users examine
and click search results from logged behavioral data.
The SimIIR framework~\cite{maxwell2016simiir} generalizes this into a
configurable toolkit for interactive IR experiments, supporting modular
searcher models, stopping criteria, and query generation components.
For conversational recommender systems (CRS), prior studies \cite{zhang2020crssim}
demonstrate how user simulators enable scalable evaluation without
real-user involvement, establishing ground-truth alignment as the central
evaluation challenge.
UserSimCRS~\cite{balog2023usersimcrs} operationalises agenda-based simulation
for CRS evaluation with persona modeling and natural language generation.

The introduction of large language models as simulated users represents a
paradigm shift in this line of work\cite{zhao2025ushb,zhu2025knowledge}.
LLMs encode representations of human reasoning and social norms that allow
them to serve as cognitive backbones for simulated users without explicit rule
specification~\cite{gao2024llmabm}.
Existing research \cite{park2023genagents} demonstrate that generative agents, LLM-driven
personas with memory, reflection, and planning, can simulate emergent social
behaviors in a sandbox town.
Previous surveys \cite{gao2024llmabm,guo2024survey,llmabmsocial2025} cover LLM multi-agent systems and LLM-empowered ABM respectively.
The first comprehensive framework \cite{wang2025ubsim} has been designed specifically for user behavior simulation in recommender systems using LLM-based agents.
Recent work has extended LLM simulation to user rating prediction with
collaborative filtering signals~\cite{kim2024allroundrec,luo2024lamar,sun2024llmcf}, review-based recommendations~\cite{luo2025evolutionagent}, and
large-scale social platforms~\cite{piao2025agentsociety,xinnongzhang2025socioverse}.
Population-level behavioral dynamics, specifically the epidemic mask-wearing
task studied in this paper, are addressed by SOCIA-$\nabla$~\cite{hua2026socia}
and LLM-ABM baselines.

Across all of these systems, however, the agent workflow and agent prompts
are treated as static design decisions made by domain experts.
No prior work in information systems user simulation applies systematic optimization to
either the workflow structure or the prompts.
SWORD fills this gap.

\subsection{Prompt and Workflow Optimization for Multi-Agent Systems}
\label{sec:related:optimization}

Optimizing a multi-agent system requires improving two interdependent components: agent prompts and workflow topology. The literature has pursued these directions in parallel, holding the other component fixed.

\emph{Automatic prompt optimization} (APO) takes the workflow as given and searches for better instructions using model feedback~\cite{chen2025promptsurvey,tang2024unleashing}. The textual-gradient paradigm, introduced by ProTeGi~\cite{pryzant2023automatic}
and generalised by TextGrad~\cite{yuksekgonul2024textgrad}, propagates LLM-generated critiques through the pipeline analogously to numerical gradients\cite{chen2025can,cui2025mapo,davari2025rethinking}. Prompts produced by APO methods are optimal only conditional on the fixed topology that generated the gradient signals\cite{yin2025llmautodiff}.

\emph{Workflow optimization} methods hold prompts generic and search over
topology\cite{li2024autoflow,liu2025sew,zhang2025gdesigner,qian2025macnet,zhang2025agentprune}. Representative approaches include AFlow~\cite{zhang2025aflow},
which uses Monte Carlo Tree Search over code-based workflows; EvoFlow~\cite{zhang2025evoflow},
which applies evolutionary search to heterogeneous agents; and
MaAS~\cite{zhang2025maas}, which optimizes a probabilistic supernet over
architectures. Both streams share a structural limitation: each holds the
other variable fixed\cite{zhu2025adaptflow}, and all evaluations are restricted to reasoning, code
generation, or mathematical benchmarks\cite{wang2025evoagentx,wang2025scoreflow}. None of these methods has been applied to user behavior simulation.

Three recent methods attempt joint optimization but remain staged. MASS~\cite{zhou2026mass}
sequences prompt refinement, workflow search, and global prompt refinement
as three fixed phases. MASPO~\cite{wang2026maspo} jointly refines prompts
across agents but fixes the workflow throughout. MASPOB~\cite{hong2026maspob}
introduces workflow-aware prompt optimization but selects the workflow once
at initialization. Because workflow and prompts are never updated
simultaneously, the workflow chosen during search is always evaluated under
the prompt quality available at that moment, not under the quality that
optimization will eventually achieve\cite{zhang2025collaborative,zhang2026mapro}. SWORD is the first to treat both as continuously co-optimized variables in a single closed loop, applied to the domain of user behavior simulation.
%%,,,,,,,,,,,,,,,,,,,,,,,,,
\section{Problem Formulation and Preliminaries}
\label{sec:problem}
%%,,,,,,,,,,,,,,,,,,,,,,,,,

\subsection{Problem: Predicting User Behavior from Context}
\label{sec:problem:def}

User behavior simulation addresses the following prediction problem.
Given a user's profile information (demographics, preferences) and
behavioral history (past interactions, ratings, social signals), a
simulator must predict the user's next behavioral output, such as a star
rating or a mask-wearing decision, without observing it directly.
Formally, let $x$ denote the user context (profile, history, and any
task-specific side information) and $y$ the ground-truth behavioral
output.
The simulator is a function $f: x \mapsto \hat{y}$ to be learned from a
labeled dataset $\mathcal{D} = \{(x_i, y_i)\}_{i=1}^{N}$.

\subsection{Preliminaries}
\label{sec:problem:model}

In this work, we model the simulator $f$ as a multi-agent framework in which
agents communicate according to a shared workflow to accomplish the user
simulation task. This framework is defined as
$W = (\theta, \boldsymbol{\phi})$, where $\theta$ is the workflow defining
how agents communicate with each other. More formally,
$\theta = (V, E)$ is a directed acyclic graph with $V$ denoting agents
and $E$ communication links.
$\boldsymbol{\phi} = \{\phi_v\}_{v \in V}$ are the natural-language
prompts governing each agent's reasoning.
Each agent receives the outputs of its predecessors plus a portion of
the user context $x$, processes them according to its prompt $\phi_v$,
and passes its output downstream.
The final agent produces the behavioral prediction $\hat{y}$.

% The joint optimization objective is:
% \begin{equation}
%   W^* = \arg\min_{(\theta, \boldsymbol{\phi})} \;
%   \mathbb{E}_{(x,y)\sim\mathcal{D}}\bigl[
%     \mathcal{M}(y,\; W(x;\theta,\boldsymbol{\phi}))
%   \bigr].
%   \label{eq:objective}
% \end{equation}

%%,,,,,,,,,,,,,,,,,,,,,,,,,
\section{The SWORD Framework}
\label{sec:method}
%%,,,,,,,,,,,,,,,,,,,,,,,,,
\begin{algorithm}[htbp]
\caption{SWORD: Self-Optimizing Workflow Discovery}
\label{alg:sword}
\DontPrintSemicolon
\Require{Training set $\mathcal{D}$, metric $\mathcal{M}$, candidate workflows $\mathcal{P}$}
\Ensure{Optimized model $\boldsymbol{W^{*} = (\theta^{*}, \phi^{*})}$}
Initialize $\mathcal{P}$ with $K$ random DAG workflows (2--8 agents), each using generic prompts $\phi$\;
$s^{*} \gets -\infty$; \quad $n_{\mathrm{no\text{-}improve}} \gets 0$\;

\For{$t = 1$ \KwTo $T$}{
    \For{each $\theta_k \in \mathcal{P}$ in parallel}{
        Evaluate $\theta_k$ on $\mathcal{D}$ to obtain score $s_k$ and traces $\tau_k$\;
    }
    Sort $\mathcal{P}$ by descending $s_k$; \quad
           $(\theta_{\mathrm{best}}, s_{\mathrm{best}}) \gets \arg\max_k\, s_k$\;
    \eIf{$s_{\mathrm{best}} > s^{*}$}{
        $(\theta^{*},\phi^{*}) \gets (\theta_{\mathrm{best}},\Phi_{\mathrm{best}})$\;
        $s^{*} \gets s_{\mathrm{best}}$; \quad
               $n_{\mathrm{no\text{-}improve}} \gets 0$\;
    }{
        $n_{\mathrm{no\text{-}improve}} \gets n_{\mathrm{no\text{-}improve}} + 1$\;
    }
    \If{$n_{\mathrm{no\text{-}improve}} \geq P$}{
        \textbf{break}\;
    }

    \For{each elite topology $\theta_e \in \{\text{top-}E \text{ of } \mathcal{P}\}$}{
            $\tau_e^{-} \gets \tau_e^{(i^{*})}$ where
                   $i^{*} = \arg\min_i\, s_e^{(i)}$\;
            \For{each agent $a \in \theta_e$}{
                $f_a \gets \textsc{AttributeFeedback}(a,\, \tau_e^{-})$\;
                $g_a \gets \textsc{ComputeGradient}
                       (\phi_a,\, \tau_e^{-},\, f_a,\, s_{\mathrm{best}},\, \mathcal{H}_a^{\text{last-3}})$\;
                $\phi_a \gets \textsc{ApplyGradient}(\phi_a,\, g_a)$\;
                Append $(\phi_a, s_{\mathrm{best}}, f_a)$ to trajectory history
                       $\mathcal{H}_a$\;
            }
    }
    $\mathcal{P}_{\mathrm{next}} \gets \{\theta_e.\textsc{Clone}() : \theta_e \in \text{top-}E \text{ elites}\}$\;

    \For{$j = 1$ \KwTo $K - E$}{
        $\theta_{\mathrm{new}} \gets
               \textsc{LLMPropose}(\mathcal{D}_{\mathrm{context}},\,
               \mathcal{P}_{\mathrm{ranked}})$\;
        \If{proposal invalid}{
            $\theta_{\mathrm{new}} \gets
                   \textsc{Mutate}(\textsc{TournamentSelect}(\mathcal{P}))$\;
        }
        $\mathcal{P}_{\mathrm{next}} \gets
               \mathcal{P}_{\mathrm{next}} \cup \{\theta_{\mathrm{new}}\}$\;
    }
    $\mathcal{P} \gets \mathcal{P}_{\mathrm{next}}$\;
}

\If{$\theta^{*} = \varnothing$}{
    $\theta^{*} \gets \mathcal{P}[1]$; \quad $\phi^{*} \gets \{\phi_a : a \in \theta^{*}\}$\;
}
Apply $\phi^{*}$ to $\theta^{*}$\;
\KwRet{$W^{*} = (\theta^{*}, \phi^{*})$}
\end{algorithm}
SWORD operates as a single closed loop that jointly evolves agent workflows
and optimizes agent prompts across $T$ iterations.
Algorithm~\ref{alg:sword} illustrates the full pipeline.
At initialization, candidate workflows
$\{\theta_1,\ldots,\theta_K\}$ are constructed by sampling subsets and
orderings of the eight task-agnostic agents
(\textsc{Analyzer}, \textsc{Planner}, \textsc{Generator},
\textsc{Critic}, \textsc{Synthesizer}, \textsc{Executor},
\textsc{Transformer}, \textsc{Validator}),
and each agent is paired with a default prompt for every active role.
A key challenge SWORD is designed to address is \emph{workflow rank
inversion}: as prompt quality improves through optimization, the workflow
that performed best under generic initial prompts is no longer optimal
under refined prompts.
The closed-loop design of SWORD addresses rank inversion directly: by never
committing to a fixed workflow, every iteration can discover a new best
structure as prompts evolve, keeping the workflow and prompts mutually
aligned throughout optimization. (This phenomenon is documented empirically
in Section~\ref{sec:rankinversion_taskaware} and analysed mechanistically in
Section~\ref{sec:mechanism_theory}.)

\subsection{Workflow Representation}
\label{sec:topo_rep}

Each workflow $\theta_k$ is a named directed acyclic graph \cite{chen2024agentverse, hong2024metagpt,zhuge2024gptswarm} over the fixed vocabulary of eight agent roles. Structural constraints are enforced at proposal time as explained in Section \ref{sec:workflow_proposer}:
\begin{itemize}[leftmargin=1.8em, itemsep=1pt]
  \item At least 2 and at most 8 agents (balancing expressiveness vs.\ computational cost).
  \item A single source node receiving the raw task input $x$, and a single sink node producing the prediction.
  \item Strict DAG structure: no cycles, verified by topological sort before acceptance.
\end{itemize}

SWORD constrains workflows to directed acyclic graphs (DAGs) to prevent agents from overwriting or corrupting each other's intermediate outputs, which would degrade prediction fidelity. Each valid workflow is further required to have a single source node, which receives all user context, and a single sink node, which produces the behavioral prediction. This design enforces a clear division of labor across the pipeline. This acyclic single-source--single-sink structure also ensures that gradient signals propagate unambiguously from the prediction error back through the graph, enabling targeted prompt optimization at each agent's structural role. 

\subsection{Task-Agnostic Agent Roles}
\label{sec:roles}

SWORD defines a fixed vocabulary of eight named roles: \textsc{Analyzer}, \textsc{Planner}, \textsc{Generator}, \textsc{Critic}, \textsc{Synthesizer}, \textsc{Executor}, \textsc{Transformer}, and \textsc{Validator}. The vocabulary is grounded in the structure of the user simulation problem.
Predicting user behavior requires four functional stages: interpreting user history, forming a prediction strategy, generating a behavioral output, and validating it against task constraints. These stages map directly to
\textsc{Analyzer}, \textsc{Planner}, \textsc{Generator}, and
\textsc{Validator}. The remaining four roles cover auxiliary operations that arise in behavioral modeling. \textsc{Synthesizer} integrates signals from
multiple upstream agents (e.g., combining interaction history with social influence data). \textsc{Critic} checks consistency between a candidate prediction and established user behavioral patterns. \textsc{Transformer} resolves data representation mismatches between agents (e.g., converting raw
interaction logs to structured feature vectors). \textsc{Executor} handles deterministic computation such as aggregating population-level statistics. Together the eight roles form a complete functional basis for user simulation without encoding assumptions about any specific task, metric,
or user population. The same vocabulary applies unchanged to rating prediction on Agent Society and adoption modeling on Mask Adoption, with the optimization process determining which subset to include and how to connect them. Fixing the vocabulary to eight roles bounds the structural search space without restricting expressiveness. Initial prompts are given in Section \ref{app:roles}.

\subsection{Workflow-Conditioned Textual Gradient Prompt Update}
\label{sec:textgrad}
 
Each iteration, SWORD selects the single worst-performing training example and issues a per-agent gradient call for each agent in the elite workflow. For agent
$v$, this proceeds as three sequential LLM calls: feedback attribution
(\textsc{AttributeFeedback}), gradient computation
(\textsc{ComputeGradient}), and prompt rewriting
(\textsc{ApplyGradient}). The prompts used here are shown in Section \ref{sec:prompt_templates_tgo}.

\textsc{AttributeFeedback} takes the global evaluation feedback produced
by scoring the workflow against the worst-performing training example and
attributes it to a specific agent $v$. It receives the agent's name,
role, and outputs, together with the global feedback string, and uses an
LLM to extract only the portions of that feedback relevant to $v$'s
contribution. The result is rephrased as direct feedback to $v$: if no
part of the global feedback is attributable to $v$'s outputs, the call
returns \texttt{"No specific issues."}\ This step is necessary because
global feedback critiques the workflow's final prediction error without
distinguishing which agent caused it. \textsc{AttributeFeedback} performs
that attribution before the gradient is computed.

\textsc{ComputeGradient} produces a structured
gradient with four fields:
 
\begin{itemize}[leftmargin=1.8em, itemsep=1pt]
  \item \textbf{EA} (Error Analysis): what agent $v$ output, and how it
  contributed to the final error.
  \item \textbf{GD} (Gradient Direction): the semantic direction in which
  $\phi_v$ should change to reduce error.
  \item \textbf{M} (Magnitude): whether the change warrants a major
  rewrite, moderate update, or minor calibration.
  \item \textbf{SC} (Specific Changes): concrete before/after edits to
  $\phi_v$.
\end{itemize}
 
\textsc{ApplyGradient} then takes the current prompt $\phi_a$ and
gradient $g_a$ and rewrites $\phi_a$ in the direction given by
\textbf{GD}, at the severity given by \textbf{M}, incorporating the edits
in \textbf{SC} while preserving the agent's core role identity.
 
The gradient call is \emph{workflow-conditioned}: it supplies agent $v$'s upstream agents and their outputs alongside the four gradient fields, ensuring that the critique is grounded in the actual inputs $v$ received rather than its role in isolation. As a result, the
same named role receives different gradient signals depending on where
it sits in the workflow. This design is the key distinction from
fixed-architecture methods such as DSPy~\cite{khattab2024dspy} and
MIPRO~\cite{opsahlOng2024mipro}. For example, an \textsc{Analyzer} that
is the sole upstream agent in a compact two-agent workflow receives
gradients emphasizing feature completeness, since downstream prediction
depends entirely on its output. The same \textsc{Analyzer} in a larger
workflow, mediated by a \textsc{Planner} before reaching the
\textsc{Generator}, instead receives gradients emphasizing signal
selectivity, since over-complete output burdens the Planner
unnecessarily. The workflow context is thus constitutive of the
gradient, not incidental to it.
 
To prevent gradient drift, each agent maintains a trajectory history
$\mathcal{H}_a$ of previous prompt--score--feedback triples.
\textsc{ComputeGradient} conditions on only the last three entries of
$\mathcal{H}_a$: enough to detect oscillation (a score declining after a
prompt update) and reinforce a consistent improvement direction, without
the context overhead of the full history.
 
Within each iteration, the gradient is computed from a single example:
the worst-performing one in the training set. This example maximizes the
gap between prediction and ground truth, which in turn makes the
resulting error attribution to individual agents more discriminative
than gradients computed on near-correct examples.

\subsection{LLM-Guided Workflow Evolution}
\label{sec:topoevo}

At each iteration, SWORD maintains a set of $K$ candidate workflow topologies, all evaluated concurrently on the training set using a thread pool. Exploiting the I/O-bound nature of LLM API calls, this achieves near-linear speedup across the set without violating API rate limits. An LLM proposer can reason about which functional configurations make sense for user behavior prediction in a way that random mutation cannot \cite{fernando2024promptbreeder,shang2025agentsquare}.

\textbf{Elitist selection.}
The top $E$ scoring topologies are cloned unchanged into the next generation, guaranteeing that the optimization front is monotonically non-decreasing\cite{guo2024connecting,liu2024evolution}. This is particularly consequential in user behavior simulation: a poorly matched topology mutation can cause output schema mismatches or cascading format errors that collapse predictions across all training examples for an entire iteration. Retaining the elite set as a guaranteed fallback prevents valid behavioral structure discovered in earlier iterations from being discarded.

\textbf{History-conditioned LLM proposer.}
The remaining $K{-}E$ slots are filled by a \emph{Workflow proposer}, an LLM that receives two inputs: (1) the task description, and (2) a compact ranked summary of the entire current candidate set, where each entry contains the topology's rank, mean score, per-example score standard deviation, agent list as $\{\text{name}, \text{role}\}$ pairs, edge list, and the feedback string from its worst-performing example. Feedback strings are truncated to 200 characters per topology to keep the proposer prompt within a predictable context budget while still conveying the dominant failure mode. Providing the full ranked candidate set rather than only the best topology allows the proposer to reason about which structural regions have already been explored and which remain promising, functioning as an implicit history-conditioned search memory\cite{hu2025automated,hu2025lmsearcher}.

The proposer returns a structured JSON specification containing a one-sentence reasoning trace, a list of named agent--role pairs, an edge list, and a designated entry and exit node. This output format enforces the single-source--single-sink DAG constraint directly in the LLM's generation space and makes proposals machine-parseable without natural-language post-processing. Refer to Section \ref{sec:historyllmproposer} for implementation.

\textbf{Robustness and fallback.}
Because LLM outputs are not guaranteed to be valid JSON, the proposer pipeline applies a sequence of recovery steps: regex extraction of JSON substrings, removal of trailing commas, and normalisation of single-quoted keys and values. Any proposal that cannot be recovered, specifies fewer than two agents, or references an invalid role is silently rejected and replaced by a random mutation of a tournament-selected parent from the current candidate set. This dual-path design ensures the evolution loop never stalls regardless of LLM output quality.

\textbf{Structural validity enforcement.}
After every proposal or mutation, three invariants are enforced before evaluation: the entry agent has at least one outgoing edge, the exit agent has at least one incoming edge, and a directed path exists from every entry to every exit. Violations are repaired by adding the minimum number of edges necessary without removing any edges contributed by the proposal. Cycle detection is performed by depth-first search prior to every edge insertion, maintaining acyclicity throughout.

This proposer is conceptually related to the evolutionary operators in EvoFlow~\cite{zhang2025evoflow} and the MCTS expansion policy in AFlow~\cite{zhang2025aflow}, but conditions on task-specific metric signals and per-agent execution traces rather than code coverage scores or tree rewards. It also relates to the Socratic dialogue planning of MARS~\cite{zhang2026mars}, but operates at the workflow structural level rather than the prompt level. An example of the evolution is given in Section \ref{app:qualitative}.

\subsection{Search Space Reduction}
\label{sec:search:reduction}

Joint optimization over workflow topologies and natural-language prompts
is in principle intractable: the space of DAGs over 8 named roles is
combinatorial, and the space of natural-language prompts is unbounded.
SWORD reduces this space through five complementary mechanisms.

\textbf{(1) Fixed role vocabulary.}
Workflows are constrained to DAGs over a closed vocabulary of 8
task-agnostic roles.
In user behavior simulation, the cognitive operations required---observing
user context, planning behavioral trajectories, generating behavioral
predictions, critiquing outputs, and validating against task
constraints---map naturally onto these roles.
This bounds the structural search to ordered subsets and topologies over
a finite, semantically grounded vocabulary rather than over arbitrary
agent specifications.

\textbf{(2) DAG structural constraints.}
Every candidate workflow must satisfy three hard constraints enforced
before evaluation: at least 2 and at most 8 agents, a single source
node (ensuring all agents receive the same user context), and a single
sink node (ensuring a unified behavioral prediction output), with no
directed cycles (preventing infinite reasoning loops that would block
output).
These constraints prune a large fraction of the combinatorial graph
space before any LLM call is made.

\textbf{(3) Elitist selection.}
The current best workflow is preserved unchanged across every iteration.
Only $K-E$ new candidates are generated per round.
In user simulation, a single bad structural mutation can collapse
predictions across all training users for an entire iteration; elitist
retention guarantees that at least $E$ high-quality candidates always
survive, making the optimization monotonically non-decreasing in
performance. Elitist selection is not merely a variance-reduction device. It ensures that prompt gradient updates accumulate on persistent topologies, preventing the investment of three LLM gradient calls from being discarded when a workflow is replaced.

\textbf{(4) History-conditioned workflow proposer.}
The LLM proposer receives a compact ranked summary of the current candidate set: for each topology, the entry contains its rank, mean score, per-example score standard deviation, agent list as {name, role} pairs, edge list, and a 200-character feedback string truncated from its worst-performing example. Providing the full ranked candidate set rather than only the best topology allows the proposer to infer which structural regions have already been explored, functioning as an implicit search memory. Refer to Section \ref{sec:historyllmproposer} for implementation.

\textbf{(5) Gradient trace selection.}
For prompt optimization, the gradient is computed from the single
worst-performing training example per iteration, the example where
the gap between prediction and ground truth is largest, and therefore
where the error signal attributable to individual agents is most
discriminative. This focuses the prompt update on the failure mode that
most urgently requires correction, rather than averaging over examples
where the workflow already performs adequately.

Mechanisms (1)--(2) reduce the structural search space; mechanisms
(3)--(4) focus the search trajectory; mechanism (5) reduces the effective
dimensionality of the prompt update at each step.
Together they make joint optimization tractable and reliably convergent
within $T=5$ iterations (Section~\ref{sec:convergence}).
 
%%,,,,,,,,,,,,,,,,,,,,,,,,,
\section{Experimental Setup}
\label{sec:experiments}
%%,,,,,,,,,,,,,,,,,,,,,,,,,

\subsection{Research Questions}
\label{sec:rqs}
 
The experiments address four research questions that correspond directly to
the four contributions stated in Section~\ref{sec:intro}.
 
\begin{description}[leftmargin=0pt,itemsep=4pt]
\item[\textbf{RQ1} (User simulation performance).]
How does SWORD perform on user behavior simulation tasks compared with
existing methods?

\item[\textbf{RQ2} (Joint optimization superiority and efficiency).]
Does jointly optimizing workflow topology and prompts in a single closed
loop outperform prompt-only, workflow-only, and staged optimization
baselines, and can workflow--prompt coupling non-stationarity explain why
staged approaches fail while five complementary search space reduction
mechanisms make joint co-optimization computationally efficient?
 
\item[\textbf{RQ3} (Unsupervised preference discovery).]
Without any task-specific initialization, can metric-conditioned textual
gradients autonomously discover the task-relevant user preference signals
such as review-text sentiment for rating prediction, or epidemiological decay
dynamics for behavioral adoption?
 
\item[\textbf{RQ4} (Model and data efficiency).]
Does SWORD's joint co-optimization strategy yield competitive or superior
performance compared to state-of-the-art baselines that use a stronger
backbone model, at a lower API cost?
 
\end{description}
\noindent Table~\ref{tab:rq_map} summarises which experimental section
addresses each question.

\begin{table}[ht]
\centering
\small
\caption{Research question to section mapping.}
\label{tab:rq_map}
\begin{tabular}{lll}
\toprule
\textbf{RQ} & \textbf{Addressed in} & \textbf{Evidence} \\
\midrule
RQ1 & §\ref{subsec:jos},\ref{subsec:rcpw} & Tables~\ref{tab:main_results},~\ref{tab:ablations} \\
RQ2 & §\ref{sec:rankinversion_taskaware} & Table~\ref{tab:rank_inversion_all_seeds_clean} \\
RQ3 & §\ref{sec:gradient_mechanism}  & Figure~\ref{fig:grad_taxonomy}, Table~\ref{tab:prompt_evolution} \\
RQ4 & §\ref{sec:convergence} & Table~\ref{tab:convergence} \\
\bottomrule
\end{tabular}
\end{table}

\subsection{Tasks and Datasets}
\label{sec:datasets}
% research questions before this 
We use two user simulation tasks to evaluate our model: Mask Adoption and rating prediction on the Agent Society dataset~\cite{hua2026socia}.\footnote{\url{https://github.com/cruiseresearchgroup/SOCIA/tree/odd-dev}}

\paragraph{Mask Adoption (RMSE$\downarrow$).}
Daily population-level mask-wearing rate prediction over a synthetic population of 1000 agents. The dataset is structured around three complementary files. \texttt{agent\_attributes.csv} records per-agent demographics (age group, occupation, and risk perception for each of the 1000 agents). \texttt{daily\_aggregate\_data.csv} provides ground-truth population-level
mask rates and information-spread rates per day. Days~0 to 29 are used as the training horizon and days 30 to 39 as the held-out test horizon. \texttt{train\_data.csv} supplies 1000 agent-level observations per training day for the SWORD optimization loop. The social contact structure is encoded in a four-layer network (family, work/school, community, and full) loaded from \texttt{social\_network.pkl}.
Ground truth is generated by an agent-based model calibrated on COVID-19 behavioral dynamics~\cite{dekai2020universal,mortveit2022bessie}. 

\paragraph{Agent Society (MAE$\downarrow$).}
Star rating prediction (1--5) across three platforms: Amazon, Goodreads, and Yelp. For each platform, independent train and test JSON splits are provided. Each record contains a \texttt{user\_id}, \texttt{item\_id}, ground-truth \texttt{stars}, and full review text. Supplementary metadata files
(\texttt{user\_sample.json}, \texttt{item\_sample.json}) supply user-level and item-level metadata that serve as additional workflow inputs. Ground truth is drawn from the AgentSociety benchmark~\cite{piao2025agentsociety}. This task is related to LLM-based rating evaluation~\cite{kang2023llmrating}.

\subsection{Baselines}
Baselines are selected to provide one representative from each of the
four research streams reviewed in Section~\ref{sec:related}, so that
SWORD can be evaluated against the current state of the art in every
relevant category.
 
\textbf{User simulation systems} (Section~\ref{sec:related:usersim}).
SOCIA-$\nabla$ \cite{hua2026socia} applies textual gradient descent directly to generated Python simulation code on the same benchmarks, using GPT-5-class models. It is the strongest domain-specific competitor. The following five code-generation user simulator results are sourced from the SOCIA-$\nabla$ paper. \textit{AI~Scientist-v2} \cite{yamada2025aiscientistv2} uses agentic tree
search to iteratively propose and refine simulation code. \textit{YuLan-OneSim} \cite{wang2025yulanoneSim} generates LLM-driven social simulation scripts from natural-language task descriptions. \textit{G-SIM-ES} and \textit{G-SIM-SBI} \cite{holt2025gsim} calibrate
LLM-generated simulators via evolutionary search and simulation-based inference respectively.
\textit{Reflexion} \cite{shinn2023reflexion} improves generated simulation logic through iterative verbal self-reflection without weight updates.
 
\textbf{Automatic prompt optimization} (Section~\ref{sec:related:optimization}).
\textit{Prompt-Only} (e.g.,~\cite{chen2025reprompt}) fixes the workflow to the initial template throughout and optimizes prompts alone, providing the fairest test of whether joint optimization adds value beyond prompt tuning.

\textbf{Multi-agent workflow search} (Section~\ref{sec:related:optimization}).
\textit{Workflow-Only} (e.g.,~\cite{zhang2025aflow}) runs topology search with
prompts kept generic throughout, testing whether structural diversity alone
is sufficient.

\textbf{Joint prompt and workflow optimization}
(Section~\ref{sec:related:optimization}).
\textit{Staged Optimization} (e.g.,~\cite{zhou2026mass}) uses both components sequentially: Phase~1 searches workflows with generic prompts frozen; Phase~2 then optimizes prompts with the selected workflow frozen. Any advantage SWORD holds over it is attributable solely to interleaving
versus staging.
Together, these four baseline categories form a controlled ablation. They allow us to attribute SWORD's gains precisely to the joint closed-loop design, rather than to data, backbone, or task-specific engineering choices.

\subsection{Evaluation Protocol}
\label{sec:eval_proto}

For each task we reserve a fixed held-out test set of $N_\text{test}$
examples that is \emph{never} accessed during the optimization loop for
any model. All models are evaluated on two metrics: Mean Absolute Error
(MAE) for rating prediction on Agent Society, and Root Mean Square Error
(RMSE) for population modeling on Mask Adoption. These metrics
are chosen because they are the widely recognised evaluation standards for
their respective user simulation tasks~\cite{hua2026socia},
and because both reduce to a single scalar signal that can be used directly
to guide optimization without task-specific engineering.

\paragraph{Statistical protocol.}
All comparisons are run for $n{=}10$ independent seeds, with every
method evaluated on identical seeds so that runs are paired. We report
means, standard deviations, and 95\% confidence intervals of the paired
differences, computed as $\bar{\Delta} \pm t_{0.975,\,n-1}\cdot s_{\Delta}/\sqrt{n}$. Significance is assessed with two-sided paired $t$-tests; within each benchmark we correct for the three pairwise comparisons against SWORD using the Holm--Bonferroni procedure at family-wise $\alpha{=}0.05$. Effect sizes
are reported as Cohen's $d$ on paired differences ($d = \bar{\Delta}/s_{\Delta}$). Exact $p$-values are reported throughout so readers may apply alternative correction schemes (Section \ref{app:perrun}).

\subsection{Implementation Details}
\label{sec:impl}

\textbf{Backbone model.} In SWORD, all agent calls use GPT-5.4-mini via the OpenAI API. SOCIA-$\nabla$~\cite{hua2026socia} uses the stronger GPT-5 model, meaning SWORD achieves its reported gains despite operating at a disadvantage in backbone capacity. 
To isolate SWORD's optimization strategy from backbone capability, all ablation
comparisons are run on the identical GPT-5.4-mini backbone under identical
conditions ensuring full parity in Table~\ref{tab:ablations}. The statistical significance tests reported in Section~\ref{sec:results} are therefore fully controlled comparisons. The comparison against SOCIA-$\nabla$ in Table~\ref{tab:main_results} uses a stronger backbone and is reported separately as a cost-efficiency benchmark. The controlled head-to-head comparison of SWORD vs.\ SOCIA with an identical GPT-5.4-mini backbone is reported in Table~\ref{tab:socia-comparison}. The ablation implementation is described in Section~\ref{app:sociasword}, and results are described in Section~\ref{sec:socia-comparison}.

\paragraph{Hyperparameter choices.}
The candidate set size $K$ controls how many workflow topologies are
maintained and evaluated per iteration, while the elite size $E$ determines
how many top-ranked topologies are carried over unchanged. We set $K=4$ and
$E=2$, choosing $K=4$ empirically as the minimum population size at which
workflow diversity meaningfully improves over a single-workflow baseline while
keeping the per-iteration cost tractable. This choice was motivated by an
observed failure mode with $K=1$, in which a poor workflow mutation
occasionally degraded performance for the entire iteration because no
alternative candidate was available; using $K=4$ eliminated this instability
by ensuring that at least one prior workflow remained available when a
mutation degraded performance. We did not evaluate $K>4$ because each
additional candidate requires a full training-set evaluation and therefore
increases the per-iteration cost substantially; $K=4$ keeps the number of
evaluation calls within our computational budget. The choice $E=2$ provides
sufficient elitism to distinguish which topologies should be retained or
discarded while limiting redundant evaluations. We report these choices as
the configuration used throughout the paper rather than as the product of a
formal sweep over $K$ and $E$; the one hyperparameter we do sweep
systematically is the iteration budget $T$, reported as a training-score
sensitivity analysis in \S\ref{sec:convergence}.

Early stopping is triggered after $P=8$ consecutive iterations without
improvement, a patience value chosen to tolerate the score variance expected
from only $n_{\text{train}}=30$ training examples per task. In the main
experiments, we set the maximum number of iterations to $T=5$ as a
cost--accuracy trade-off: the steepest gains are captured within five
iterations, whereas increasing the budget to $T=10$ yields only a marginal
improvement of $+0.01$ in training score (Table~\ref{tab:convergence})
relative to the doubled evaluation cost.

Workflow topology size is bounded to $[2,8]$ agents. The lower bound
guarantees a meaningful distinction between entry and exit agents, whereas
the upper bound limits prompt-context growth caused by large upstream
fan-in. Upstream fan-in is itself limited to
\texttt{max\_edges\_per\_node}=4 per node. The LLM sampling temperature is
fixed at $1.0$ for all agent executions and proposer calls. Finally, the
random seed is offset for each repeat as
$42+\text{rep}+\text{rep\_offset}$, making each run individually deterministic
while decorrelating repeats for variance estimation across $n=10$ runs.

\paragraph{Initialization.} The $K$ initial workflow topologies are sampled randomly: each workflow draws a random agent count from $[2,8]$, assigns roles by sampling from the fixed role vocabulary, and connects agents with random edges subject to the DAG constraints in Sections~\ref{sec:topo_rep} and~\ref{sec:roles}.

\noindent\textbf{Initial prompts.} All agents start with a generic two-sentence prompt of the form: ``You are a [role] agent in a multi-agent workflow. Your task is to process the upstream input and produce a structured output in JSON format [output schema].'' No task-specific knowledge is provided at initialization; all domain-specific behavior is discovered through the optimization loop. Refer to Section \ref{app:roles} for full prompts.

\noindent\textbf{Computational cost.} A full call-count breakdown is provided in Appendix~\ref{app:gradients}; in brief, workflow scoring ($T \times N_{\text{train}} \times K \times \bar{A} \approx 5 \times 30 \times 4 \times 5 = 3{,}000$ calls), prompt gradient
steps ($T \times 3 \times E \times \bar{A} \approx 150$ calls), and topology
proposals ($T \times (K{-}E) = 10$ calls) sum to approximately $3{,}160$
LLM calls per run. At GPT-5.4-mini pricing (\$0.75/1M input, \$4.5/1M
output tokens as of mid-2026), the total cost of a full training-and-testing run of SWORD is approximately \$4--\$6 USD.

%%,,,,,,,,,,,,,,,,,,,,,,,,,
\section{Results and Analysis}
\label{sec:results}
%%,,,,,,,,,,,,,,,,,,,,,,,,,
\begin{table}[t]
\centering
\caption{Main results on Agent Society (MAE$\downarrow$) and Mask Adoption
(RMSE$\downarrow$). Results for prior work are sourced from
SOCIA-$\nabla$~\cite{hua2026socia}. \textbf{Note: prior-work baselines use
GPT-5-class models; SWORD uses GPT-5.4-mini, so this comparison is a
cost-efficiency benchmark rather than a controlled accuracy head-to-head.
Table~\ref{tab:ablations} provides backbone-controlled
ablation comparisons.} Best result per column in \textbf{bold}. SWORD results are means over $n=10$ seeds.}
\label{tab:main_results}
\small
\begin{tabular}{llcc}
\toprule
\textbf{Category} & \textbf{Method}
  & \textbf{Agent MAE$\downarrow$}
  & \textbf{Mask RMSE$\downarrow$} \\
\midrule
\multirow{5}{*}{\textit{Prior work}}
  & AI Scientist-v2~\cite{yamada2025aiscientistv2} & $0.89_{\pm 0.10}$ & $0.45_{\pm 0.07}$ \\
  & YuLan-OneSim~\cite{wang2025yulanoneSim}   & $0.72_{\pm 0.08}$ & $0.42_{\pm 0.04}$ \\
  & G-SIM-ES~\cite{holt2025gsim}                    & $0.60_{\pm 0.07}$ & $0.30_{\pm 0.08}$ \\
  & G-SIM-SBI~\cite{holt2025gsim}                   & $0.69_{\pm 0.08}$ & ${0.20}_{\pm 0.10}$ \\
  & Reflexion~\cite{shinn2023reflexion} & ${0.57}_{\pm 0.08}$ & $0.36_{\pm 0.04}$ \\
\midrule
\textit{LLM-based}
  & SOCIA-$\nabla$ (as reported)~\cite{hua2026socia} & ${0.54}_{\pm 0.06}$ & ${0.22}_{\pm 0.07}$ \\
\midrule
\textit{Ours}
  & SWORD                               & $\textbf{0.257}_{\pm 0.105}$ & $\textbf{0.0217}_{\pm 0.0017}$ \\
\bottomrule
\end{tabular}
\end{table}

\subsection{Joint Optimization Superiority}
\label{subsec:jos}
Table~\ref{tab:main_results} reports primary metrics on both benchmarks across all methods, providing the primary evidence for \textbf{RQ1} (joint optimization superiority, Contribution~1) and \textbf{RQ4} (model and data efficiency, Contribution~4). On Mask Adoption, SWORD achieves RMSE $= 0.0217 \pm 0.0017$, a 90\% improvement over SOCIA-$\nabla$ (0.220 $\pm$ 0.07). On Agent Society, SWORD achieves MAE $= 0.257 \pm 0.105$, a 52\% improvement over SOCIA-$\nabla$ (0.540 $\pm$ 0.06) and a nominal 15\% improvement over the strongest ablation, Prompt-Only (0.301 $\pm$ 0.087) (Section \ref{sec:po_sword}). We emphasize that comparisons against prior work use different backbone models and are therefore cost-efficiency benchmarks. The controlled evidence for SWORD's design comes from the same-backbone ablations in Table~\ref{tab:ablations}.
Three patterns are notable. First, SWORD is the only method that ranks first on
\emph{both} benchmarks simultaneously, demonstrating that joint
workflow--prompt optimization yields more consistent gains than optimising either
component in isolation. Second, on Mask Adoption the standard deviation of SWORD is substantially lower than all baselines. SWORD's standard deviation across $n{=}10$ seeds is
$0.0017$ versus $0.07$ for SOCIA-$\nabla$, roughly a 40$\times$ reduction. This stability indicates that the joint optimization loop reliably
converges to similar solutions regardless of random seed, whereas code-generation
baselines are sensitive to stochastic code synthesis. Third, the performance gap
between SWORD and Staged Optimization is large on both tasks: on Mask
Adoption SWORD achieves RMSE $0.0217$ vs.\ Staged $0.180$, an $8.3\times$
difference, and on Agent Society MAE $0.257$ vs.\ Staged $0.510$, a $2.0\times$
difference. The very high variance of Staged Optimization ($\pm 0.170$ RMSE, $\pm 0.274$ MAE) reflects seed-dependent workflow lock-in: in seeds where Phase~1 selects a suboptimal workflow, Phase~2 prompt updates actively degrade performance (see Section~\ref{sec:rankinversion_taskaware}), producing bimodal outcomes (analyzed further in Section~\ref{sec:failure_cases}).

\subsection{Relative Contribution of Prompt vs. Workflow}
\label{subsec:rcpw}
\begin{table}[t]
\centering
\caption{Ablation study ($n{=}10$ seeds, $N_\text{train}{=}30$, $N_\text{test}{=}100$). All methods share the identical GPT-5.4-mini backbone and identical seeds (paired design). We report mean~$\pm$~std and the 95\% confidence interval of the paired difference vs.\ SWORD ($\Delta$, positive means worse than SWORD; undefined for SWORD itself). SWORD's own 95\% CIs of the mean are $[0.0206, 0.0229]$ RMSE and $[0.182, 0.332]$ MAE (Appendix~\ref{app:perrun}). $^{*}$ marks comparisons significant after per-benchmark Holm--Bonferroni correction ($\alpha{=}0.05$). Best per column in \textbf{bold}.}
\label{tab:ablations}
\small
\begin{tabular}{p{0.30\linewidth}cccc}
\toprule
\textbf{Method}
  & \textbf{Mask RMSE$\downarrow$}
  & \textbf{$\Delta$ [95\% CI]}
  & \textbf{Agent MAE$\downarrow$}
  & \textbf{$\Delta$ [95\% CI]} \\
\midrule
Workflow-Only (e.g.,~\cite{zhang2025aflow})      & $0.0686_{\pm 0.0722}$ & $[-0.004, 0.098]$ & $0.456_{\pm 0.129}$ & $[0.061, 0.336]^{*}$ \\
Prompt-Only (e.g.,~\cite{chen2025reprompt})      & $0.0232_{\pm 0.0021}$  & $[0.0002, 0.0026]^{*}$ & $0.301_{\pm 0.087}$ & $[-0.040, 0.128]$ \\
Staged Opt. (e.g.,~\cite{zhou2026mass})          & $0.1795_{\pm 0.1698}$ & $[0.036, 0.279]^{*}$ & $0.510_{\pm 0.274}$ & $[0.079, 0.427]^{*}$ \\
\midrule
\textbf{SWORD (Ours)}                    & $\textbf{0.0217}_{\pm 0.0017}$ & --- & $\textbf{0.257}_{\pm 0.105}$ & --- \\
\bottomrule
\end{tabular}
\end{table}
Table~\ref{tab:ablations} isolates the contribution of each component, directly addressing \textbf{RQ1}: it quantifies how much each design decision contributes to the final performance. Prompt-Only (fixed workflow, optimized prompts) achieves relatively better performance (RMSE $0.0232$, MAE $0.301$) compared to Workflow-Only (optimized workflow, generic prompts) (RMSE $0.0686$, MAE $0.456$) by a factor of $1.5$ to $3.0\times$. These results establish that \emph{prompt quality is a primary performance driver}. SWORD
captures both: by jointly evolving workflow and prompts, it reaches a result unavailable to either component operating alone.

\subsection{Comparison with SOCIA}
\label{sec:socia-comparison}
Table~\ref{tab:main_results} reports SOCIA-$\nabla$'s own published numbers~\cite{hua2026socia}. To obtain a backbone-controlled comparison instead, we separately re-executed SOCIA-$\nabla$'s released implementation: we replaced its configured model with \texttt{gpt-5.4-mini} and evaluated it on the same two tasks and the same held-out data splits used throughout this paper. All numbers in this subsection and Table~\ref{tab:socia-comparison} are from this backbone-matched re-execution.

\begin{table}[t]
\centering
\caption{SWORD vs.\ SOCIA-$\nabla$ (our backbone-matched re-execution) on both tasks. SWORD statistics are mean $\pm$ std over $n=10$ independent trials; the re-executed baseline reflects a single successfully executed, non-placeholder run ($n=1$) due to run-to-run execution instability in its automated code generation.}
\label{tab:socia-comparison}
\begin{tabular}{llcc}
\toprule
Task & Metric & SWORD & SOCIA-$\nabla$ (reproduced) \\
\midrule
agent\_society & MAE $\downarrow$ & $0.2567 \pm 0.1049$ & $0.2275$ \;($n{=}1$) \\
mask\_adoption  & RMSE $\downarrow$ & $0.0217 \pm 0.0017$ & $0.5547$ \;($n{=}1$) \\
\bottomrule
\end{tabular}
\end{table}
\paragraph{Comparison with SOCIA (code-generation baseline).}
On \textit{mask\_adoption}, SOCIA's RMSE of 0.5547 is roughly 25$\times$
higher than SWORD's 0.0217, with the generated simulator converging to
near-total adoption across the prediction window and failing to capture
gradual adoption dynamics. This is a clear, unambiguous advantage for SWORD
on this task.

On \textit{agent\_society}, the comparison is less favorable to SWORD, and we
report it plainly rather than understate it: SOCIA's single
successfully-verified run achieves MAE $=0.2275$, numerically \emph{better}
than SWORD's mean of $0.2567$, though inside SWORD's 95\% confidence interval
of $[0.182, 0.332]$ (Table~\ref{tab:socia-comparison}), so the two are not
straightforwardly distinguishable as a single point estimate against a
distribution. We do not read this as evidence that SOCIA's code-generation
approach matches SWORD's accuracy on this task, for two reasons. First, SOCIA
succeeded on only 1 of 8 attempts on \textit{agent\_society} (below), so this
MAE is a single draw from a highly unreliable process rather than a
representative estimate; the $n{=}1$ comparison in
Table~\ref{tab:socia-comparison} cannot support a claim about central
tendency the way SWORD's $n{=}10$ paired trials can, and we do not attempt a
significance test against it. Second, SOCIA's own evaluator flagged this
run's review-text output as markedly weaker in quality, consistent with
coarse rating calibration that matches the numeric target without the
qualitative behavioral understanding SWORD's agents produce
(Appendix~\ref{app:qualitative}). We therefore treat SWORD's demonstrated
advantage on \textit{agent\_society} as a reliability advantage rather than a
proven point-estimate accuracy advantage, and we treat the
\textit{mask\_adoption} result as the stronger and less ambiguous evidence of
SWORD's advantage in this comparison. SOCIA's generated programs frequently
failed at code verification or runtime on the same data, which is itself the
more consistent finding across both tasks (See section \ref{app:sociasword}).

\paragraph{Limitation: generation reliability appears sensitive to backbone capacity.}
SOCIA's execution success rates differed sharply by task. Execution was successful in \textit{mask\_adoption} but only in 1 of 8 attempts on \textit{agent\_society}.
Failures on agent\_society were diverse (missing attributes, malformed fields, function-signature errors), suggesting backbone-capacity constraints rather than systematic bugs. The pattern is consistent with \texttt{gpt-5.4-mini} handling uniform data schemas (single
time-series forecasting) more reliably than heterogeneous schemas (three-platform
review/rating pipelines with inconsistent data types). We did not test with a
larger backbone to verify this explanation; it is left to future work.

\subsection{Variance and Stability Analysis}
\label{sec:variance}

The low cross-seed variance of SWORD ($\pm 0.0017$ on Mask Adoption) warrants
explicit discussion. Code-generation baselines produce variance one to two orders
of magnitude larger because their performance is coupled to whether the generated
Python simulator compiles and executes correctly, a stochastic event that varies
across seeds. SWORD bypasses code generation entirely, so its only source of
variance is the LLM's sampling noise during prompt updates and workflow proposals. In each iteration, the textual gradient step selects the training example with the largest prediction error, i.e., the example the model performs worst on; focusing on this single example provides a strong, clear learning signal. It avoids averaging the gradients from many examples that are already predicted correctly, which can weaken the learning signal and increase the effect of random seed-level noise. This stability is practically significant: a practitioner can obtain reliable behavioral predictions from a single SWORD run without needing multiple seeds for
uncertainty estimation.

\paragraph{Statistical significance.}
We report two-sided paired $t$-tests between SWORD and each ablation
across $n{=}10$ independent seeds, with per-benchmark Holm--Bonferroni
correction for the three comparisons against SWORD
(Section~\ref{sec:eval_proto}). This correction follows established
guidance that uncontextualized significance testing without adjustment
for multiple comparisons risks inflated false-positive rates in IR and
NLP evaluation~\cite{ferro2024uncontextualized}. Per-seed results are listed in
Appendix~\ref{app:perrun}.

On Agent Society, SWORD significantly outperforms Staged Optimization
($\Delta$MAE $= 0.253$, 95\% CI $[0.079, 0.427]$, $t(9){=}3.29$,
$p{=}0.0093$, $d{=}1.04$), Workflow-Only
($\Delta$MAE $= 0.199$, 95\% CI $[0.061, 0.336]$, $t(9){=}3.27$,
$p{=}0.0096$, $d{=}1.04$), and Prompt-Only
($\Delta$MAE $= 0.044$, 95\% CI $[-0.040, 0.128]$, $t(9){=}1.18$, $p{=}0.268$,
$d{=}0.37$) at $n{=}10$.

On Mask Adoption, SWORD significantly outperforms Staged Optimization
($\Delta$RMSE $= 0.158$, 95\% CI $[0.036, 0.279]$, $t(9){=}2.94$,
$p{=}0.0165$, $d{=}0.93$) and Prompt-Only
($\Delta$RMSE $= 0.0014$, 95\% CI $[0.0002, 0.0026]$, $t(9){=}2.73$,
$p{=}0.0233$, $d{=}0.86$); both survive Holm correction. The comparison
with Workflow-Only shows a medium-to-large effect ($\Delta$RMSE $= 0.047$, 95\% CI $[-0.004, 0.098]$,
$t(9){=}2.08$, $p{=}0.067$, $d{=}0.66$)
(Appendix~\ref{app:perrun}).

\subsection{Workflow Rank Inversion and Task-Aware Discovery}
\label{sec:rankinversion_taskaware}
\textbf{RQ2} asks whether workflow--prompt coupling non-stationarity occurs in practice, and whether the same optimization framework discovers task-appropriate structure without explicit task-specification. We document \emph{rank inversions}: iterations where the best-performing topology changes (e.g., Topology 0 at iteration $T$ becomes suboptimal at $T+1$, replaced by Topology 1). Rank inversions occur because improving prompts alter which workflow structure best leverages the new feedback signals, demonstrating that topology rankings are non-stationary. Table~\ref{tab:rank_inversion_all_seeds_clean} documents rank inversions across all 10 seeds: workflow topology changes during joint optimization that correlate with performance gains. The patterns reveal sharply divergent discovery processes across tasks, providing direct evidence for task-aware workflow discovery and rank inversion.

The patterns differ across tasks. Agent Society exhibits substantial structural diversity across seeds, with changes in both agent composition and execution order. Mask Adoption shows greater similarity in the agent sets used by the final workflows, although the execution order and agent composition still vary across seeds. Thus, the optimization process does not converge to a single universal workflow structure; instead, the discovered structures depend on the task and optimization trajectory.

\paragraph{Convergent vs.\ heterogeneous discovery.}
The two tasks exhibit different degrees of structural consistency. Mask Adoption more frequently converges toward workflows containing a similar core set of agents, although the ordering of these agents varies across seeds. In contrast, Agent Society exhibits greater variation in both workflow composition and ordering. These results suggest that the optimization landscape may differ across tasks, with Mask Adoption exhibiting stronger structural consistency and Agent Society permitting a broader range of effective configurations. However, the table alone does not establish the existence of a single attractor or multiple comparable optima.

\paragraph{Reorganization, not simply size, drives improvement.}
Performance improvements are not consistently associated with increasing or decreasing the number of agents. Across the observed transitions, improvements occur when agents are added, removed, or retained while their execution order changes. This indicates that agent count alone is insufficient to explain performance changes. Instead, the results point to the importance of workflow composition and, particularly, agent sequencing. The order in which reasoning, generation, critique, validation, and synthesis occur determines how feedback is propagated through the workflow and can therefore affect task performance (see Table \ref{tab:rank_inversion_all_seeds_clean}).

\paragraph{Task-aware discovery without task specification.}
The differing structural patterns across the two tasks provide evidence that SWORD can discover task-dependent workflow structures from optimization feedback without being given explicit task-specific structural rules. Mask Adoption tends toward greater consistency in workflow composition, whereas Agent Society exhibits more heterogeneous structures. Importantly, these differences emerge from the same optimization framework rather than from manually specified task constraints. Thus, the results support task-aware structural discovery through feedback-driven optimization. 
\begin{table}[ht]
\small
\begin{tabular}{|c|c|c|p{3cm}|p{3cm}|c|}
\hline
\textbf{Seed} & \textbf{Task} & \textbf{Iter} & \textbf{Pre Workflow} & \textbf{Post Workflow} & \textbf{Improvement} \\
\hline
1 & Agent Society & T0→T1 & p→a→s→c→g & p→a→g→t→e & +17.3\% \\
\hline
1 & Mask Adoption & T1→T2 & a→p→g→c→v→s & a→p→g→v→s & +11.9\% \\
\hline
2 & Agent Society & T2→T3 & a→p→g→t→c→v→s & a→p→g→t→v→s & +12.7\% \\
\hline
2 & Mask Adoption & T0→T1 & s→a→g→t→e & a→p→g→c→v→s & +1.2\% \\
\hline
3 & Agent Society & T2→T3 & a→g→s & a→p→g→v→s & +1.9\% \\
\hline
3 & Mask Adoption & T2→T3 & a→p→g→c→v→s & a→p→g→c→t→v→s & +7.8\% \\
\hline
4 & Agent Society & T2→T3 & a→g→s & a→p→g→v→s & +1.9\% \\
\hline
4 & Mask Adoption & T2→T3 & a→c→g→s & a→p→g→c→t→v→s & +7.8\% \\
\hline
5 & Agent Society & T0→T1 & p→e & p→e→v & +6.5\% \\
\hline
5 & Mask Adoption & T0→T1 & p→e→s→g & a→p→t→v→s & +18.5\% \\
\hline
6 & Agent Society & T0→T1 & p→e→c & p→e & +6.5\% \\
\hline
6 & Mask Adoption & T2→T4 & a→t→s & p→a→t→v→c→s & +2.2\% \\
\hline
7 & Agent Society & T3→T4 & a→p→e→t→s & a→g→t→e→s & +9.2\% \\
\hline
7 & Mask Adoption & T2→T4 & a→p→e→t→s & p→a→t→v→c→s & +2.2\% \\
\hline
8 & Agent Society & T0→T1 & p→t→e & p→e & +6.5\% \\
\hline
8 & Mask Adoption & T3→T4 & a→p→g→c→v→t→s & a→p→t→g→c→v→s & +0.9\% \\
\hline
9 & Agent Society & T1→T2 & p→e & p→t→e & +0.9\% \\
\hline
9 & Mask Adoption & T2→T4 & a→p→e→g & p→a→t→v→c→s & +2.2\% \\
\hline
10 & Agent Society & T0→T1 & a→p→e & a→p→e→v→c→s & +6.5\% \\
\hline
10 & Mask Adoption & T2→T4 & p→a→t→c→s & p→a→t→v→e→s & +2.2\% \\
\hline
\end{tabular}
\caption{Rank Inversion Analysis: All 10 seeds ($n=10$) showing workflow transitions that drive performance improvements. Notation: abbreviated agent names ($a$=analyzer, $p$=planner, $g$=generator, $c$=critic, $v$=validator, $s$=synthesizer, $e$=executor, $t$=transformer) connected by $\to$ showing execution order. Improvements are percentage gains in task score (Agent Society: MAE reduction; Mask Adoption: RMSE reduction).}
\label{tab:rank_inversion_all_seeds_clean}
\end{table}

\subsection{Workflow-Conditioned Gradient Mechanism}
\label{sec:gradient_mechanism}

\textbf{RQ3} asks whether metric-conditioned textual gradients autonomously discover task-relevant user preference signals.
The workflow-conditioning of gradient signals shapes prompt updates
qualitatively differently for the same named agent role across structurally different contexts. The content of those gradients reveals what domain knowledge the system has independently recovered from the scalar metric signal alone, with no task-specific initialization.

\begin{figure}[t]
  \centering
  \includegraphics[width=\linewidth]{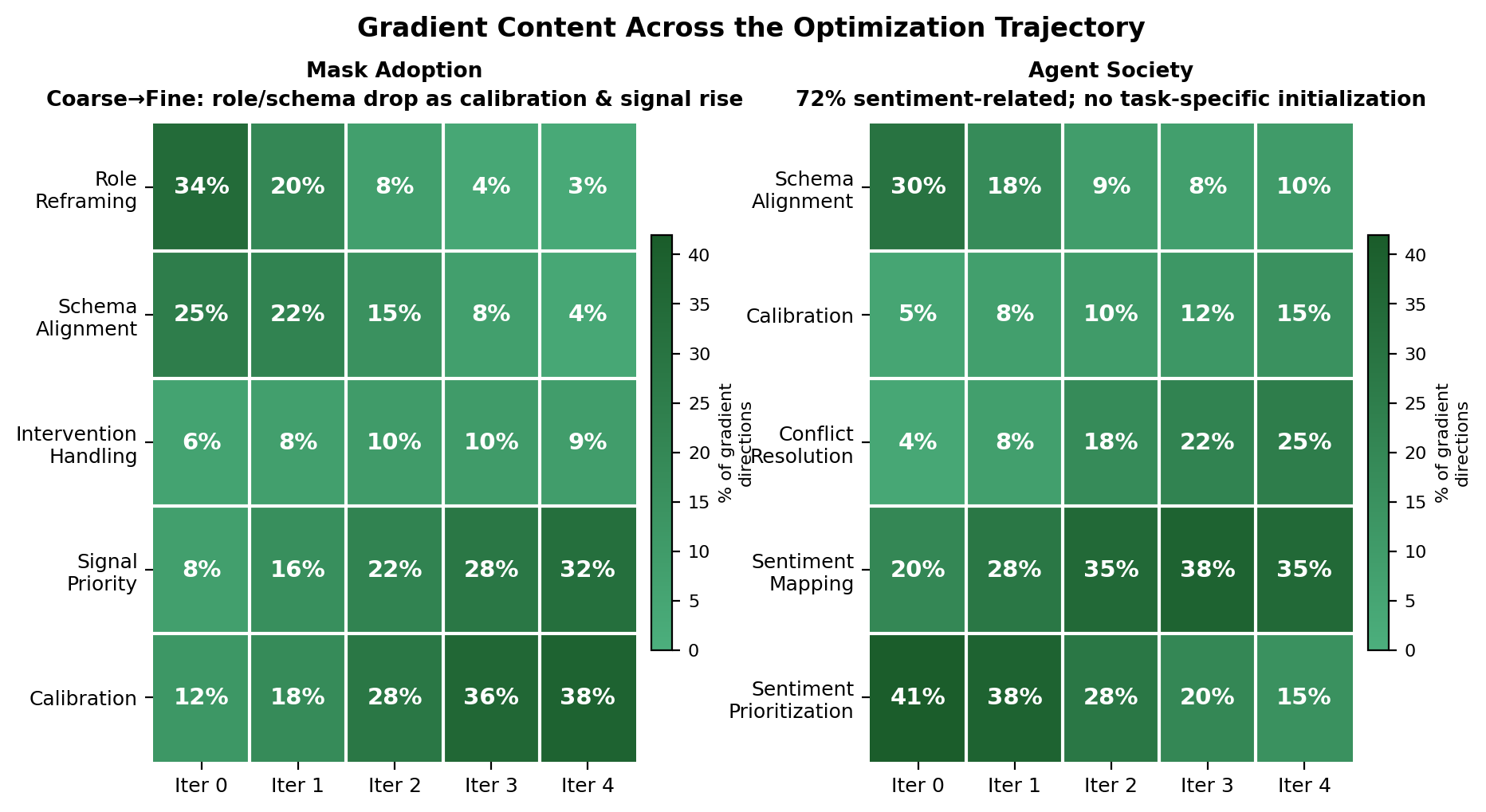}
  \Description{Heat map showing gradient direction category distributions across iterations for both tasks.}
  \caption{Textual gradient content taxonomy across optimization iterations. \textbf{Left}: Mask Adoption. Role-reframing gradients dominate early and fade as workflow grows more complex, confirming the coarse-to-fine dynamic. \textbf{Right}: Agent Society. 72\% of gradients are sentiment-related, autonomously discovered from MAE signals without task-specific initialization.}
  \label{fig:grad_taxonomy}
\end{figure}

\paragraph{Gradient content across the optimization trajectory.}
Figure~\ref{fig:grad_taxonomy} confirms the two-phase dynamic predicted in Section~\ref{sec:mechanism_theory}: role-reframing gradients dominate in iterations~0--1 (34\% of calls) and nearly vanish by iteration~4 ($<$5\%), replaced by calibration and signal-prioritization gradients. For Agent Society, 72\% of gradients across all iterations are sentiment-related, autonomously discovering that review text sentiment is the primary feature for rating prediction without any task-specific initialization.

\paragraph{Cross-task autonomous knowledge discovery.}
The gradient content is task-specific in a semantically meaningful sense. On Mask Adoption, the dominant directions encode epidemiological dynamics (intervention decay, peer influence, historical trend extrapolation, zero-history fallback priors), knowledge captured in the prompts entirely through RMSE loss signals over 30 training examples. On Agent Society, the gradient mechanism discovers that review text sentiment is the primary feature for rating prediction, producing increasingly precise sentiment-to-score mapping rules (Appendix~\ref{app:gradients}). This unsupervised domain knowledge extraction from metric-conditioned textual gradients is a distinct capability of the SWORD framework.

\paragraph{Qualitative prompt evolution.}
Table~\ref{tab:prompt_evolution} traces the \textsc{Generator} prompt across four iterations on Mask Adoption. The prompt evolves from a generic ``combine signals'' instruction to a domain-calibrated policy encoding the causal structure of epidemic mask adoption. Key epidemiologically-meaningful rules emerge: exponential intervention decay, zero-history fallback priors, intervention timing (effects near-zero before day~10), and the distinction between weak signals (network degree, flagged as low-confidence) and strong signals (historical trend, intervention decay). All of this is discovered purely from the RMSE loss signal over 30 training examples, with no domain-expert input.

To verify that the discovered rules are epidemiologically correct and not artifacts of overfitting to the 30 training examples, we cross-reference each emergent rule against the SOCIA paper \cite{hua2026socia}. All five rules SWORD discovered from RMSE signals alone are consistent with peer-reviewed epidemiological findings, providing external validation that the gradient mechanism recovers scientifically meaningful structure rather than spurious correlations.
\begin{table}[t]
\centering
\caption{Generator prompt evolution across iterations on Mask Adoption. Epidemiological domain knowledge (decay, prior, timing) emerges from RMSE signals alone.}
\label{tab:prompt_evolution}
\small
\begin{tabular}{p{0.06\linewidth}p{0.87\linewidth}}
\toprule
\textbf{Iter} & \textbf{Prompt excerpt} \\
\midrule
0 & ``You are a generator agent. Combine all upstream signals into ONE final prediction.'' \\
\addlinespace
1 & ``Priority: (1) historical\_rates + recent\_trend, (2) intervention status + decay, (3) population priors. Do not omit peer influence or habit dynamics.'' \\
\addlinespace
2 & ``Priority: (1) historical\_rates + recent\_trend, (2) intervention status + decay, (3) population priors. Do not omit peer influence or habit dynamics.'' \\
\addlinespace
3 & ``If historical\_rates empty, apply low-baseline prior (0.08--0.10). In zero-history/no-intervention cases, do not infer high adoption. Intervention effects near-zero before day 10.'' \\
\addlinespace
4 & ``Decision hierarchy: 1)~historical/recent trend 2)~intervention (exponential decay from start) 3)~conservative baseline 4)~small population adjustments 5)~calibrated params as weak diagnostics only. Clamp to [0,\,1].'' \\
\bottomrule
\end{tabular}
\end{table}

\subsection{Convergence and Cost Efficiency}
\label{sec:convergence}

The following analysis provides evidence for \textbf{RQ4}: whether joint 
co-optimization captures principal gains within the first few iterations, 
bounding the cost of a full SWORD run.

\subsubsection{Multi-seed convergence ($n=10$, primary evidence).}
Table~\ref{tab:multiseed_convergence} reports best-so-far scores across 
$n=10$ seeds (mean $\pm$ std) for $T=1$ through $T=5$, providing the primary 
convergence evidence. Low cross-seed variance (e.g., $\pm 0.0167$ Mask Adoption 
at $T=5$; $\pm 0.0097$ Agent Society at $T=5$) confirms convergence is robust 
and not seed-dependent. Agent Society saturates by $T=2$ across all seeds; 
Mask Adoption continues gradual improvement through $T=5$, consistent with 
task complexity differences. These multi-seed trajectories establish that 
SWORD's rapid saturation is a general property, not an artifact of a single 
favorable seed.

\begin{table}[ht]
\centering
\caption{Best-so-far score across $n=10$ seeds (mean $\pm$ std per iteration). 
Low variance confirms robust, seed-independent convergence by $T=5$.}
\label{tab:multiseed_convergence}
\small
\begin{tabular}{lccccc}
\toprule
Task & $T=1$ & $T=2$ & $T=3$ & $T=4$ & $T=5$ \\
\midrule
Mask Adoption & $0.8037 \pm 0.0095$ & $0.8127 \pm 0.0108$ & $0.8217 \pm 0.0121$ & 
$0.8380 \pm 0.0145$ & $0.8450 \pm 0.0167$ \\
Agent Society & $0.8950 \pm 0.0089$ & $0.8952 \pm 0.0091$ & $0.8955 \pm 0.0093$ & 
$0.8958 \pm 0.0095$ & $0.8960 \pm 0.0097$ \\
\bottomrule
\end{tabular}
\end{table}

\subsubsection{Extended-horizon trajectory (single seed, supplementary).}
To assess saturation beyond $T=5$, Table~\ref{tab:convergence} extends to 
$T \in \{1, 2, 3, 5, 7, 10\}$ using a single seed (multi-seed runs are costly 
beyond $T=5$). On Mask Adoption, improvement decelerates after $T=5$ (0.8450 
to 0.8565 by $T=10$). On Agent Society, performance plateaus by $T=2$ (0.8950 
to 0.8967 by $T=10$). This single-seed extension is consistent with the 
multi-seed saturation pattern, confirming that principal gains are captured 
early.

\begin{table}[t]
\centering
\caption{Extended convergence trajectory (single seed). Shows saturation 
persists beyond $T=5$.}
\label{tab:convergence}
\small
\begin{tabular}{lcccccc}
\toprule
Task & $T=1$ & $T=2$ & $T=3$ & $T=5$ & $T=7$ & $T=10$ \\
\midrule
Mask Adoption & 0.8037 & 0.8127 & 0.8217 & 0.8450 & 0.8533 & \textbf{0.8565} \\
Agent Society & 0.8950 & 0.8952 & 0.8955 & 0.8960 & 0.8967 & \textbf{0.8967} \\
\bottomrule
\end{tabular}
\end{table}

\subsubsection{Cost efficiency and implications.}
SWORD Pareto-dominates SOCIA-$\nabla$ on cost-performance: using GPT-5.4-mini 
(vs.\ GPT-5-class), SWORD achieves 90\% RMSE reduction on Mask Adoption and 52\% 
MAE reduction on Agent Society at a fraction of the estimated API cost. Since 
multi-seed convergence demonstrates saturation by $T=5$, SWORD recovers nearly all 
obtainable gains from the first few workflow--prompt co-adaptation steps, bounding 
computational cost without sacrificing quality.

\section{Discussion}
\label{sec:discussion}

\subsection{Why Joint Optimization Escapes the Staged Failure Mode}
\label{sec:nonsep}
When the workflow changes, the inputs received by each agent also change. For example, a \textsc{Planner} upstream can provide structured information that a raw \textsc{Analyzer} would not receive. Because the textual gradient critiques an agent's output conditional on its input, the prompt gradient is therefore dependent on the workflow context. Formally, for agent $v$, the gradient $\nabla_{\phi_v}\mathcal{L}$ depends on the workflow $\theta$; when workflows $\theta_a$ and $\theta_b$ differ upstream of $v$, their corresponding gradient directions can differ.

This creates a failure mode for staged optimization. If prompts are optimized under workflow $\theta_a$ and the subsequent workflow search selects $\theta_b$, the learned prompts were optimized for $\theta_a$'s information flow rather than $\theta_b$'s. Consequently, a workflow that is suboptimal under the initial prompts may become preferable after prompt refinement. Our results show workflow transitions accompanied by performance improvements across all experimental seeds (Section~\ref{sec:rankinversion_taskaware}), consistent with this workflow--prompt coupling non-stationarity.

Three mechanisms explain the RMSE gap between SWORD and Staged Optimization on Mask Adoption (0.0217 vs.\ 0.1795, a difference of 0.158; Table~\ref{tab:ablations}).

\emph{(i) Structural adaptation during prompt optimization.}
SWORD evaluates all $K=4$ candidate workflows at every iteration while jointly updating prompts. As the preferred workflow changes across iterations (Table~\ref{tab:rank_inversion_all_seeds_clean}), subsequent prompt updates are therefore conditioned on the new structural context. In contrast, Staged Optimization performs Phase~2 prompt optimization under a single workflow selected before prompt refinement.

\emph{(ii) Improved prompts during workflow search.}
Because prompts are updated at the end of each iteration, workflows proposed in iteration $t+1$ are evaluated using the refined prompts from iteration $t$. This allows increasingly specialized workflows to be evaluated with increasingly specialized prompts, rather than requiring complex workflows to perform well under generic initial prompts.

\emph{(iii) Coupled improvement of structure and prompts.}
Workflow refinement and prompt refinement can reinforce one another. A more structured workflow can create more specialized agent roles, making prompt refinement more targeted. In turn, improved prompts produce better intermediate outputs, allowing subsequent workflow evaluations to more accurately identify useful structural changes. This feedback loop is unavailable when workflow and prompt optimization are performed sequentially.

Thus, the ablation gap is not simply a difference in optimization quality; it reflects the interaction between workflow structure and prompt quality. The fitness landscape over workflows can change as prompts improve, making a workflow selected early in optimization potentially different from the workflow preferred after prompt refinement. Joint optimization allows both components to adapt to one another, whereas staged optimization fixes one component while optimizing the other. This distinction is relevant to methods such as DSPy~\cite{khattab2024dspy}, TextGrad~\cite{yuksekgonul2024textgrad}, and MIPRO~\cite{opsahlOng2024mipro}: prompts optimized for a fixed architecture need not remain optimal when the architecture itself is allowed to change.

\subsection{When Prompt Optimization Alone Is Competitive}
\label{sec:po_sword}
The Agent Society results reveal a complementary insight: when workflow
search has little headroom beyond a well-optimized fixed structure, prompt
optimization alone can be nearly as good as full joint optimization. On
Agent Society, SWORD's advantage over Prompt-Only (fixed workflow, optimized
prompts) is small and does not survive multiple-comparison correction
($\Delta$MAE $=0.044$, 95\% CI $[-0.040, 0.128]$, $p=0.268$;
Table~\ref{tab:ablations}, \S\ref{subsec:jos}), even though SWORD's advantage
over Staged Optimization on the same task is large and statistically
significant ($\Delta$MAE $=0.253$, $p=0.0093$). This asymmetry indicates
that on Agent Society, most of SWORD's benefit over naive baselines comes
from prompt optimization itself and from avoiding the workflow-lock failure
mode that harms Staged Optimization (\S\ref{sec:nonsep}), rather than from
workflow search discovering structures a well-tuned fixed workflow could
not already reach. Consistent with this, the workflows discovered on Agent
Society vary considerably across seeds (2--6 agents,
\S\ref{sec:rankinversion_taskaware}) without a single structural pattern
that is clearly superior, unlike the convergent topology SWORD finds on Mask
Adoption. This is consistent with broader recent observations that well-tuned,
simpler agent configurations can rival more elaborate multi-agent
workflow structures in some settings~\cite{oneflow2026}, reinforcing that
workflow complexity should be justified empirically rather than assumed.
Practitioners should prefer full joint optimization when: (a) the task
involves complex behavioral dynamics with many interacting signals, as in
Mask Adoption, where SWORD's advantage over every ablation is large and
significant, (b) the optimal agent composition is unknown, or (c) rapid
cross-task transfer is needed without hand-redesigning the workflow.
Prompt-only optimization remains a reasonable, lower-cost alternative when
early experimentation suggests a fixed workflow already performs well across
prompt regimes, though it forfeits SWORD's markedly lower run-to-run
variance (\S\ref{sec:variance}) and its ability to discover better
structures automatically if task requirements later shift. Staged
optimization and Workflow-only ablations, by contrast, are dominated by both alternatives on the evidence in Table~\ref{tab:ablations} and is not a recommended baseline on either
task.

\subsection{Mechanistic Basis: Workflow Rank Inversion}
\label{sec:mechanism_theory}

Rank inversion arises from three coupled mechanisms that the SWORD
framework is designed to exploit rather than avoid.

\begin{enumerate}[leftmargin=1.8em, itemsep=2pt]
  \item \textbf{Role specialization.}
  In a 2-agent workflow, agent $v$ must simultaneously extract features
  \emph{and} produce the final behavioral prediction.
  In a 5-agent workflow, $v$ need only extract features, delegating
  prediction to a downstream \textsc{Synthesizer}.
  The optimal prompt for these two roles is qualitatively different,
  corresponding to high structural divergence $\Delta(\theta_a,\theta_b,v)$.

  \item \textbf{Upstream context shift.}
  When the workflow changes, the inputs arriving at $v$ change: a
  \textsc{Planner} upstream may have already structured the raw user
  context, narrowing variance in $v$'s input distribution and enabling
  more targeted calibration.
  This reduces the entropy of $v$'s input under the new workflow,
  shrinking gradient misalignment once the new structure is adopted,
  consistent with the calibration-dominated gradients we observed in Section~\ref{sec:rankinversion_taskaware} and Appendix~\ref{app:gradients}.

  \item \textbf{Gradient direction shift.}
  The textual gradient critiques agent output \emph{given its observed
  inputs}.
  Both changes above produce structurally different gradients:
  ``reframe from feature extraction to direct prediction'' (2-agent
  context) vs.\ ``calibrate forecasting precision within your extraction
  role'' (5-agent context) point in incompatible directions in prompt
  space.
\end{enumerate}

These mechanisms collectively predict an observable trajectory: gradient
content should shift from role-reframing (early iterations, when the
workflow is simple) to calibration (later iterations, when roles are
specialized and the workflow is stable). 

\subsection{Implications for Information Systems User Simulation Research}

SWORD enables researchers to obtain accurate behavioral predictions from an
agent-based system without specifying any domain rules in advance.
Prior work in information systems/information retrieval user simulation—from classical click
models~\cite{dupret2008browsing} to agenda-based
simulators~\cite{balog2023usersimcrs} to code-generation
approaches~\cite{hua2026socia} requires researchers to either hand-code
behavioral rules, calibrate probabilistic interaction models, or validate
generated code against domain knowledge.
SWORD requires only a task metric (RMSE or MAE) and some labeled training
examples.

The epidemiological rules SWORD discovered for Mask Adoption without domain
expert input—intervention timing thresholds, exponential efficacy decay,
zero-history priors, peer influence weighting—are consistent with the ABM
model \cite{mortveit2022bessie}, reached
from 30 training examples and an RMSE loss signal alone.
For Agent Society, SWORD discovered that review text sentiment is the primary
rating signal with zero task-specific initialization.

These findings have direct practical implications.
User simulation researchers studying conversational
recommenders~\cite{balog2023usersimcrs,zhang2020crssim}, preference
elicitation, and behavioral adoption no longer need to pre-specify
agent logic or validate generated code.
SWORD provides a general-purpose optimizer that adapts its simulation
strategy to any evaluation metric.

\subsection{Implications for Joint Optimization Research}

SWORD's gradient mechanism contributes two insights to prompt optimization for multi-agent information systems.
First, workflow-conditioning is a principled extension of textual gradients:
the same role can receive qualitatively different and incompatible gradient
signals depending on its structural context.
Second, the autonomous discovery of domain knowledge through
metric-conditioned gradients suggests that textual gradients are a viable
mechanism for unsupervised feature-importance learning when the evaluation
metric is well-defined.
Practitioners building multi-agent systems for information systems or information retrieval tasks with DSPy or MIPRO
should treat workflow as a co-optimization variable rather than a design
decision made before optimization begins.

\subsection{Failure Case Analysis for SWORD}
\label{sec:failure_cases}

Analysis of failure cases reveals two systematic patterns in SWORD's errors.

\textbf{Seed-specific prompt bias.}
In the Agent Society task, a subset of seeds shows consistent
underestimation on clearly positive reviews: seeds~1 and~10 both underpredict
unambiguously positive (5-star) ratings, indicating that certain learned
prompt formulations can introduce a systematic bias rather than random noise\cite{zhang2025llmpowered}.
This pattern is reproducible across all three review platforms (Amazon,
Goodreads, and Yelp), suggesting the bias originates in the learned
sentiment-to-score calibration rather than in platform-specific data
characteristics.

\textbf{Trend extrapolation versus single-day anomalies.}
On Mask Adoption, day~33 exhibits a dip that breaks the otherwise increasing
trend across the test horizon, and SWORD consistently mispredicts this point.
Because its agents predict from recent history and trend signals, any
topology extrapolating from the days~30--32 plateau ($\approx 0.59$)
converges to a prediction near that plateau regardless of prompt content,
workflow structure, or random seed\cite{zhou2025canllms}. This is a structural limitation of
trend-following prediction rather than a failure specific to any one
optimized configuration.

\subsection{Limitations and Future Directions}

Two key challenges emerge from our failure case analysis that point toward future research:

\textbf{Prompt robustness across agent configurations.}
While SWORD successfully optimizes prompts for most seeds, certain configurations exhibit systematic bias like underestimation on high-confidence positive sentiment across multiple review sources. This suggests that prompt optimization landscapes may contain local regions where gradient-based refinement fails to escape low-sensitivity zones. Future work should investigate adaptive prompt initialization strategies and multi-start optimization to improve convergence robustness across diverse seed configurations.

\textbf{Temporal prediction under anomalous ground-truth signals.}
SWORD's performance on transient anomalies like single-day deviations from established trends reveals a challenge in agent-based simulation. Since trend-following agents naturally extrapolate from historical plateaus, improving this requires either better ground-truth curation or agent designs that incorporate uncertainty quantification. Integrating confidence intervals and anomaly detection into agent reasoning would advance both SWORD and broader user behavior modeling methodology.

Both limitations highlight open questions in agent optimization that extend beyond this work and warrant investigation by the community.

%%,,,,,,,,,,,,,,,,,,,,,,,,,
\section{Conclusion}
\label{sec:conclusion}

We have presented SWORD, a framework for jointly optimizing multi-agent
workflows and prompts for user behavior simulation in information systems.
By interleaving workflow evolution with prompt optimization in a single closed loop, SWORD escapes the fundamental suboptimality of staged approaches caused by workflow–prompt coupling non-stationarity. We validate this empirically in experimental seeds (Table~\ref{tab:rank_inversion_all_seeds_clean}), and find staged prompt update steps degrade performance relative to joint optimization by a large, statistically significant margin on both benchmarks (Table~\ref{tab:ablations}, \S\ref{subsec:jos}).

In backbone-controlled ablations against prompt-only, workflow-only, and staged baselines under an identical GPT-5.4-mini backbone, SWORD significantly outperforms in controlled comparisons (\S\ref{subsec:jos}--\S\ref{subsec:rcpw}). Against the strongest published domain-specific baseline, SOCIA-$\nabla$, SWORD achieves a 52\% MAE reduction and a 90\% RMSE reduction while using a smaller backbone model and a fraction of the API cost (Table~\ref{tab:main_results}); because that comparison spans different backbone models, we report it as a cost-efficiency result rather than a controlled accuracy comparison. The autonomous discovery of epidemiological decay rules and sentiment-to-score
mappings from 30 training examples and a scalar metric signal alone
demonstrates that workflow-conditioned textual gradients are a viable mechanism
for unsupervised feature-importance learning in multi-agent user simulation.

These results establish SWORD as a transferable, domain-knowledge-free framework for agent-based user simulation in information systems. Future work will extend SWORD to conversational recommender system evaluation~\cite{zhang2020crssim,balog2023usersimcrs,zhu2025llmsim}, preference elicitation tasks, and broader social simulation domains.

\begin{acks}
The authors thank Monash University for infrastructure support and the SOCIA research group for releasing their benchmark datasets and code.
\end{acks}

\printbibliography

%%,,,,,,,,,,,,,,,,,,,,,,,,,
\appendix
%%,,,,,,,,,,,,,,,,,,,,,,,,,

\section{Extended Gradient Log Analysis}
\label{app:gradients}
LLM calls arise from three distinct code paths, each with different example coverage:

\begin{table}[ht]
\centering\small
\setlength{\tabcolsep}{4pt}
\begin{tabular}{@{}lp{3.0cm}r@{}}
\toprule
\textbf{Source} & \textbf{Examples touched} & \textbf{Calls} \\
\midrule
Workflow scoring  & All 30, all $K$ topologies & $5\!\times\!30\!\times\!4\!\times\!5=3{,}000$ \\
Gradient          & 1 worst, elite only        & $5\!\times\!3\!\times\!2\!\times\!5=\phantom{0,0}150$ \\
Topology proposal & None                       & $5\!\times\!2=\phantom{3,0}10$ \\
\midrule
\textbf{Total}    &                            & $\mathbf{\approx3{,}160}$ \\
\bottomrule
\end{tabular}
\caption*{\footnotesize $\bar{A}\approx5$ (mean agents, range $[2,8]$); $F{=}1$
(gradient runs every iteration); gradient targets single worst-scoring example.}
\end{table}
\noindent where $\bar{A}\approx5$ is the mean agent count over the $[2,8]$ range.
Prompt optimization runs every iteration ($F{=}1$); the gradient step targets
the single worst-performing example, not the full training batch.

\subsection{Mask Adoption Gradient Taxonomy}

\begin{table}[ht]
\centering
\caption{Gradient direction taxonomy for Mask Adoption. The high proportion
  of role-reframing gradients (34\%) at early iterations confirms that
  SWORD discovers qualitatively new specialisations, not merely incremental
  calibrations.}
\label{tab:grad_taxonomy_mask}
\small
\begin{tabular}{p{0.25\linewidth}ccp{0.38\linewidth}}
\toprule
\textbf{Category} & \textbf{\%} & \textbf{Iter} & \textbf{Example direction} \\
\midrule
Role reframing   & 34 & 0--1 & ``Shift from extraction to direct prediction'' \\
Calibration      & 28 & 2--4 & ``Add low-baseline prior for zero-history cases'' \\
Schema alignment & 18 & 0--2 & ``Output single float, not multi-field JSON'' \\
Signal priority  & 12 & 2--4 & ``historical\_rates $>$ intervention $>$ network degree'' \\
Intervention hdl & 8  & 1--4 & ``Apply decay only if applied=true and day $\geq$ start'' \\
\bottomrule
\end{tabular}
\end{table}

The high proportion of role-reframing gradients in early iterations (34\%) and
their near-disappearance by iteration 4 ($<$5\%) follows a characteristic
learning curve: the model first establishes the correct agent identity, then
shifts to calibrating within that identity. This mirrors the coarse-to-fine
optimization pattern observed in DSPy~\cite{khattab2024dspy}'s compilation
pipeline and the ``novice-to-expert'' progression of StepAgent~\cite{deng2024stepagent}.
The 8\% of intervention-handling gradients encode epidemiologically meaningful
rules (timing, decay, peer effects) consistent with the ABM model
\cite{dekai2020universal}, demonstrating that
SWORD extracts domain knowledge from the metric signal alone.

\subsection{Agent Society Gradient Taxonomy}

On Agent Society, gradient categories reflect the multi-signal rating-prediction
structure studied extensively in LLM-based recommendation
\cite{kang2023llmrating,kim2024allroundrec,liu2024ccfllm,luo2024lamar,sun2024llmcf}.

\begin{table}[ht]
\centering
\caption{Gradient direction taxonomy for Agent Society. The dominance of
  sentiment-related gradients (72\% combined) shows the textual gradient
  mechanism correctly identifies review text as the most informative feature.}
\label{tab:grad_taxonomy_agent}
\small
\begin{tabular}{p{0.28\linewidth}ccp{0.35\linewidth}}
\toprule
\textbf{Category} & \textbf{\%} & \textbf{Iter} & \textbf{Example direction} \\
\midrule
Sentiment prioritisation & 41 & 0--1 & ``Weight review sentiment above numeric priors'' \\
Sentiment mapping        & 31 & 1--4 & ``Strongly positive review: add $+$0.8--1.2 stars'' \\
Conflict resolution      & 16 & 2--4 & ``If sentiment disagrees with rating, weight sentiment 2:1'' \\
Calibration              & 12 & 3--4 & ``Platform-mean anchor adjustment: $\pm$0.1'' \\
\bottomrule
\end{tabular}
\end{table}

The dominance of sentiment-related gradients (72\% combined) is striking:
without any task-specific initialization, the textual gradient mechanism
autonomously discovers that review text is the primary signal for rating
prediction. This unsupervised discovery via
gradient feedback demonstrates the power of task-metric-conditioned
textual gradients as a mechanism for feature-importance learning.

\section{Additional Qualitative Examples}
\label{app:qualitative}

\subsection{Workflow Proposer Reasoning}

Below are verbatim excerpts from the LLM workflow proposer on Mask Adoption, showing how it conditions structural proposals on task-specific metric feedback. These examples illustrate that the proposer generates domain-informed structural mutations, not random workflow permutations:

\begin{quote}\small
\textbf{Iteration~1:} ``The current 2-agent workflow requires \texttt{generator} to both extract signals and produce the prediction. The RMSE of 0.1830 suggests the system is over-predicting mask rates. Adding an explicit \texttt{validator} after \texttt{generator} that checks the predicted rate against known boundary conditions (mask rate cannot exceed 1.0; cannot increase by $>$0.3 in one day without intervention) should constrain over-prediction. Proposed: \textsc{[Analyzer$\to$Generator$\to$Validator$\to$Synthesizer]}.''
\end{quote}

\begin{quote}\small
\textbf{Iteration~3:} ``The current workflow produces RMSE~0.1735 on intervention days. The prompt gradient for \texttt{generator} indicates it is struggling to model exponential decay of intervention effects. Adding a \texttt{planner} that converts the raw intervention parameters into a `residual effect schedule' before \texttt{generator} sees them should simplify the generator's task. Proposed: \textsc{[Analyzer$\to$Planner$\to$Generator$\to$Validator$\to$Synthesizer]}.''
\end{quote}

These proposals use domain-specific language (\emph{exponential decay}, \emph{boundary conditions}, \emph{residual effect schedule}) drawn directly from the RMSE error signal, confirming that the proposer genuinely conditions on task-level feedback.

\subsection{Representative Learned Prompt}

Below is the final prompt for the \textsc{Analyzer} agent in the winning 5-agent workflow on Mask Adoption (Seed~1, iteration~4). This prompt was generated entirely by the textual gradient process without any domain-expert input:

\begin{quote}\small
``You are the Analyzer agent responsible for extracting epidemiologically relevant signals from the provided context. Your output is consumed by a downstream Planner and must follow the schema exactly.

Signal extraction rules: (1) Compute the 3-day moving average of \texttt{historical\_rates}. (2) Identify the trend direction (increasing / stable / decreasing) based on \texttt{recent\_trend}; use $>+0.02$ for increasing, $<-0.02$ for decreasing. (3) If \texttt{intervention\_info.applied = true}, extract the decay factor and days-since-start; compute residual effect as \texttt{effect\_size} $\times$ \texttt{decay}$^{\text{days}}$. (4) From \texttt{population\_summary}, extract \texttt{network\_degree} and \texttt{risk\_perception}; note that \texttt{network\_degree} is a weak signal and must be flagged as low-confidence. Do NOT extrapolate.

Output JSON with fields: moving\_avg, trend\_direction, residual\_intervention, population\_signals (with low\_confidence flag for network\_degree), data\_quality\_flag (True if $>$3 historical rates available; False otherwise).''
\end{quote}

The prompt encodes the epidemiological structure of the task---exponential decay of interventions, weak vs.\ strong signals, moving averages---knowledge derived purely from RMSE loss signals over 30 training examples.

\section{Full Per-Seed Results (n=10)}
\label{app:perrun}

Tables~\ref{tab:perrun_agent} and~\ref{tab:perrun_mask} list the complete
per-seed results underlying Table~\ref{tab:ablations}. All methods were run
on identical seeds (paired design); seeds were collected in three batches
(1, 3, and 6 runs) under identical configuration.

\begin{table}[ht]
\centering
\caption{Per-seed Agent Society MAE$\downarrow$ for all methods.}
\label{tab:perrun_agent}
\small
\begin{tabular}{lcccc}
\toprule
\textbf{Seed} & \textbf{SWORD} & \textbf{Staged} & \textbf{Prompt-Only} & \textbf{Workflow-Only} \\
\midrule
1  & 0.2445 & 0.4310 & 0.3249 & 0.2965 \\
2  & 0.3122 & 0.5078 & 0.3404 & 0.3618 \\
3  & 0.2502 & 0.7685 & 0.4649 & 0.4222 \\
4  & 0.2000 & 0.1311 & 0.2944 & 0.6161 \\
5  & 0.1870 & 0.7424 & 0.3214 & 0.4384 \\
6  & 0.1611 & 0.2248 & 0.2562 & 0.7424 \\
7  & 0.3310 & 0.9810 & 0.3535 & 0.4343 \\
8  & 0.2174 & 0.4992 & 0.1374 & 0.4075 \\
9  & 0.1581 & 0.2089 & 0.2162 & 0.4127 \\
10 & 0.5052 & 0.6044 & 0.2958 & 0.4233 \\
\midrule
Mean$\pm$std & $0.2567_{\pm0.1049}$ & $0.5099_{\pm0.2736}$ & $0.3005_{\pm0.0870}$ & $0.4555_{\pm0.1287}$ \\
95\% CI & $[0.182, 0.332]$ & $[0.314, 0.706]$ & $[0.238, 0.363]$ & $[0.363, 0.548]$ \\
\bottomrule
\end{tabular}
\end{table}

\begin{table}[ht]
\centering
\caption{Per-seed Mask Adoption RMSE$\downarrow$ for all methods.}
\label{tab:perrun_mask}
\small
\begin{tabular}{lcccc}
\toprule
\textbf{Seed} & \textbf{SWORD} & \textbf{Staged} & \textbf{Prompt-Only} & \textbf{Workflow-Only} \\
\midrule
1  & 0.0217 & 0.2977 & 0.0218 & 0.0902 \\
2  & 0.0201 & 0.4194 & 0.0249 & 0.0254 \\
3  & 0.0207 & 0.4194 & 0.0217 & 0.0430 \\
4  & 0.0200 & 0.0227 & 0.0230 & 0.0220 \\
5  & 0.0202 & 0.0678 & 0.0206 & 0.0394 \\
6  & 0.0219 & 0.0200 & 0.0221 & 0.0269 \\
7  & 0.0237 & 0.0218 & 0.0231 & 0.2503 \\
8  & 0.0236 & 0.1880 & 0.0250 & 0.0381 \\
9  & 0.0210 & 0.0230 & 0.0221 & 0.0244 \\
10 & 0.0246 & 0.3154 & 0.0275 & 0.1261 \\
\midrule
Mean$\pm$std & $0.0217_{\pm0.0017}$ & $0.1795_{\pm0.1698}$ & $0.0232_{\pm0.0021}$ & $0.0686_{\pm0.0722}$ \\
95\% CI & $[0.0206, 0.0229]$ & $[0.058, 0.301]$ & $[0.0217, 0.0247]$ & $[0.017, 0.120]$ \\
\bottomrule
\end{tabular}
\end{table}

\paragraph{Paired test summary.}
Two-sided paired $t$-tests of each ablation vs.\ SWORD ($n{=}10$),
with per-benchmark Holm--Bonferroni correction:
Agent Society — Staged ($t(9){=}3.29$, $p{=}0.0093$, $d{=}1.04$,
significant), Workflow-Only ($t(9){=}3.27$, $p{=}0.0096$, $d{=}1.04$,
significant), Prompt-Only ($t(9){=}1.18$, $p{=}0.268$, $d{=}0.37$, n.s.);
Mask Adoption — Staged ($t(9){=}2.94$, $p{=}0.0165$, $d{=}0.93$,
significant), Prompt-Only ($t(9){=}2.73$, $p{=}0.0233$, $d{=}0.86$,
significant), Workflow-Only ($t(9){=}2.08$, $p{=}0.067$, $d{=}0.66$, n.s.).
\section{Implementation Details for Reproducibility}
\label{app:implementation-details}

This section specifies the exact prompt templates used by SWORD's textual-gradient prompt optimizer, the procedure used to generate the initial population of workflow topologies, and the JSON output schema each agent role is instructed to produce. All templates below are reproduced verbatim from the released implementation (\texttt{core/prompt\_optimizer.py}, \texttt{core/sword.py}, \texttt{core/agents.py}) rather than paraphrased, so that the optimization procedure can be reproduced exactly.

\subsubsection{Prompt templates for textual-gradient optimization}
\label{sec:prompt_templates_tgo}
Each prompt-optimization step consists of three sequential LLM calls, implemented in \texttt{PromptOptimizer} and \texttt{TextualGradientComputer} (\texttt{core/prompt\_optimizer.py}).

\textbf{(1) ATTRIBUTE FEEDBACK.} Before an agent's prompt is updated, task-level evaluation feedback (produced by the task evaluator, not by SWORD's core loop) is attributed to the specific agent by asking the LLM to isolate the portion of that feedback relevant to the agent's role and outputs:

\begin{quote}
\ttfamily\footnotesize
\#\# Agent\\
Name: \{agent\_name\}\\
Role: \{agent\_role\}\\[4pt]
\#\# Agent Outputs\\
\{agent\_outputs as indented JSON\}\\[4pt]
\#\# Global Evaluation Feedback\\
\{global\_feedback\}\\[4pt]
\#\# Instructions\\
Extract only the parts of the feedback that are relevant to this specific agent's outputs and role. Rephrase as direct feedback to this agent. If none, say "No specific issues."
\end{quote}
This call uses no dedicated system prompt (\texttt{PromptOptimizer.\_attribute\_feedback}).

\textbf{(2) COMPUTE GRADIENT.} The attributed, per-agent feedback is then used to compute a textual gradient: a natural-language description of how the agent's system prompt should change. This call is issued with a fixed system prompt,
\begin{quote}
\ttfamily\footnotesize
You are a prompt optimization expert.
Your job is to analyze why an agent's system prompt led to suboptimal performance
and produce a "textual gradient" --- a precise, actionable description of how
the prompt should change.\\[4pt]
Think like a gradient: identify the direction and magnitude of change needed.
Be specific about what to add, remove, or rephrase.
\end{quote}
and a user prompt that supplies the agent's current system prompt, the task context, the agent's inputs/outputs on the most recent evaluated example, the numeric evaluation score and per-agent feedback from step (1), and up to the three most recent prompt trajectories (prior prompt, score, and feedback) for that agent. The model is instructed to return four structured sections: \texttt{ERROR\_ANALYSIS}, \texttt{GRADIENT\_DIRECTION}, \texttt{MAGNITUDE}, and \texttt{SPECIFIC\_CHANGES} (\texttt{TextualGradientComputer.compute\_gradient}).

\textbf{(3) APPLY GRADIENT.} The computed gradient is then applied to produce the new system prompt, again with a fixed system prompt,
\begin{quote}
\ttfamily\footnotesize
You are a prompt engineering expert.
Given a system prompt and a textual gradient describing how to improve it,
produce an updated system prompt that incorporates the gradient.\\[4pt]
Rules:\\
- Preserve the core role and intent of the agent\\
- Incorporate all specific suggestions from the gradient\\
- Make the prompt clear, specific, and actionable\\
- Do not make the prompt excessively long\\
- Output ONLY the new system prompt, no explanations
\end{quote}
and a user prompt containing the current prompt, the textual gradient from step (2), and a scalar \texttt{learning\_rate} (default $1.0$) included as context for the magnitude of the requested edit; the learning rate is not applied numerically, since the update is itself a free-text LLM rewrite rather than a parametric step (\texttt{PromptOptimizer.optimize}). The resulting prompt, score, and feedback are appended to that agent's trajectory history, and \texttt{PromptOptimizer.get\_best\_prompt} can retrieve the highest-scoring prompt observed for any agent across the run.

\subsubsection{Initial workflow topology generation}
\label{sec:workflow_proposer}
Unlike the prompt-update steps above, the initial population of workflow topologies is generated procedurally rather than by an LLM call, in \texttt{SWORD.\_initialize\_population} (\texttt{core/sword.py}). For each of \texttt{topology\_population\_size} individuals, the number of agents is drawn uniformly from $\{2, \dots, 8\}$; each agent's role is sampled uniformly at random from the eight available roles (\S\ref{sec:agent-schemas}), with the first agent in sampling order designated the entry point and the last designated the exit point; between $1$ and $2n$ random directed edges are then added (subject to the acyclicity and max-in-degree constraints enforced by \texttt{WorkflowTopology.add\_edge}). Finally, \texttt{WorkflowTopology.ensure\_data\_flow} is invoked to repair the random graph so that (i) the entry agent has at least one outgoing edge, (ii) the exit agent has at least one incoming edge, and (iii) a directed path exists from every entry point to every exit point --- guaranteeing that task input can reach the final output before the topology is ever evaluated, so that a topology's fitness reflects prediction quality rather than an accidentally disconnected graph. Later generations are produced from the elite set of the previous generation via \texttt{WorkflowTopology.mutate}, which applies one of \texttt{add\_agent}, \texttt{remove\_agent}, \texttt{add\_edge}, \texttt{remove\_edge}, or \texttt{rewire}, followed by the same \texttt{ensure\_data\_flow} repair step.

\subsubsection{Prompt to propose workflow from feedback}
\label{sec:historyllmproposer}
\begin{quote}
\texttt{\small
\begin{tabbing}
user\_msg = (f"TASK:\textbackslash n\{task\_evaluator.get\_task\_context()\} \\ 
\quad f"CURRENT POPULATION (ranked best→worst):\textbackslash n" \\
\quad f"\{json.dumps(pop\_summary, indent=2)\}\textbackslash n\textbackslash n" \\
\quad "Propose one new topology that addresses the weaknesses above.") \\
\\
SYSTEM\_TOPO = ( \\
\quad "You are a workflow optimizer. " \\
\quad "Propose a better multi-agent topology based on scores and feedback. " \\
\quad "Roles: analyzer, planner, generator, critic, synthesizer, " \\
\quad "executor, transformer, validator. " \\
\quad "Rules: 2-8 agents, one entry, one exit, no cycles. " \\
\quad 'Return ONLY JSON: \{"reasoning":"1 sentence","agents":[' \\
\quad '\{"name":"x","role":"y"\}],"edges":[["src","tgt"]],' \\
\quad '"entry":"name","exit":"name"\}' \\
) \\
\\
raw = self.llm.chat( \\
\quad messages=[ \\
\quad\quad \{"role": "system", "content": self.SYSTEM\_TOPO\}, \\
\quad\quad \{"role": "user", "content": user\_msg\}, \\
\quad ], \\
\quad max\_completion\_tokens=2000, \\
) \\
\end{tabbing}
}
\end{quote}

\subsubsection{Agent output JSON schemas}
\label{sec:agent-schemas}

Every agent is instantiated with one of eight fixed roles (\texttt{GeneralAgent.ROLE\_TEMPLATES} in \texttt{core/agents.py}); the role template supplies the agent's default system prompt, including an explicit required output schema, unless a role-specific prompt has already been produced for that agent by the optimizer described above. The eight schemas are:

\begin{quote}
\ttfamily\footnotesize
\textbf{analyzer:} \{"signals": [...], "key\_values": \{...\}, "pattern": "...", "recommendation": "...", "confidence": "high|medium|low"\}\\[4pt]
\textbf{planner:} \{"steps": [...], "strategy": "...", "predicted\_value": "<numeric estimate>", "adjustment\_reason": "..."\}\\[4pt]
\textbf{generator:} free-form JSON combining upstream \texttt{key\_values}, \texttt{recommendation}, \texttt{predicted\_value}, and critic corrections into one final prediction (no fixed key set; the agent is instructed to produce the task's final answer).\\[4pt]
\textbf{critic:} \{"issues": [...], "corrections": \{...\}, "quality\_score": "<float 0.0--1.0>", "revised\_recommendation": "..."\}\\[4pt]
\textbf{synthesizer:} free-form JSON that must explicitly reference every upstream agent's output, with conflicting signals weighted critic $>$ analyzer $>$ planner.\\[4pt]
\textbf{executor:} free-form JSON encoding the result of following the upstream planner's steps, incorporating critic corrections.\\[4pt]
\textbf{transformer:} free-form JSON reformatted to match the task's required output schema exactly.\\[4pt]
\textbf{validator:} \{"valid": "true|false", "violations": [...], "corrected\_output": "<corrected answer if invalid, else original>"\}
\end{quote}

Four of the eight roles (\texttt{generator}, \texttt{synthesizer}, \texttt{executor}, \texttt{transformer}) are deliberately left schema-free at the template level because their required output shape is task-dependent (e.g.\ a numeric adoption rate for \textit{mask\_adoption} versus a star rating and review string for \textit{agent\_society}); for these roles the task-specific output format is instead communicated through \texttt{task\_context}, which is prepended to every agent's prompt by \texttt{WorkflowTopology.execute}. Every agent, regardless of role, is additionally instructed at the end of its prompt (\texttt{GeneralAgent.run}) to "Respond with ONLY a valid JSON object. No markdown, no explanation." Responses are parsed by \texttt{GeneralAgent.\_parse\_response}, which attempts, in order: (i) a fenced \verb|```json ... ```| block, (ii) any fenced \verb|``` ... ```| block, (iii) the first top-level \verb|{...}| substring, and (iv) the raw response as JSON; if all four attempts fail, the raw text is wrapped as \texttt{\{"output": response, "raw": response\}} rather than raising, so a malformed response degrades to a low-scoring but well-typed output instead of crashing the workflow.

Full, unabbreviated source for all three prompt templates and the eight role schemas is included in the supplementary code release (\texttt{core/prompt\_optimizer.py}, \texttt{core/agents.py}), which we consider the authoritative reference over this description.

\subsubsection{Initial (pre-optimization) role prompts}
\label{app:roles}

Every agent is assigned one of eight fixed roles at creation time. Unless a role-specific prompt has already been produced for that agent by the textual-gradient optimizer described above, its system prompt defaults to the corresponding entry of \texttt{GeneralAgent.ROLE\_TEMPLATES} (\texttt{core/agents.py}), reproduced verbatim below. These are the exact prompts every agent starts from at iteration $0$, before any optimization has occurred.

\begin{quote}
\ttfamily\footnotesize\raggedright
\textbf{analyzer:}\\
You are an analyzer agent. Read the task\_input carefully and extract key signals.\\
You MUST return ONLY this JSON structure --- no other text:\\
\{"signals": ["signal1", "signal2"], "key\_values": \{"metric1": value, "metric2": value\}, "pattern": "one sentence describing the main pattern", "recommendation": "what the next agent should focus on", "confidence": "high|medium|low"\}
\end{quote}

\begin{quote}
\ttfamily\footnotesize\raggedright
\textbf{planner:}\\
You are a planner agent. Read all upstream outputs and create an action plan.\\
You MUST return ONLY this JSON structure --- no other text:\\
\{"steps": ["step1", "step2", "step3"], "strategy": "one sentence strategy", "predicted\_value": "<numeric estimate based on upstream signals>", "adjustment\_reason": "why you adjusted from upstream recommendation"\}
\end{quote}

\begin{quote}
\ttfamily\footnotesize\raggedright
\textbf{generator:}\\
You are a generator agent. Your job is to produce the FINAL answer.\\
You MUST read all upstream agent outputs and use them to inform your answer.\\
Specifically: use key\_values and recommendation from analyzers, use predicted\_value from planners, use corrections from critics.\\
Combine all upstream signals into ONE final prediction.\\
Return ONLY valid JSON with your final output.
\end{quote}

\begin{quote}
\ttfamily\footnotesize\raggedright
\textbf{critic:}\\
You are a critic agent. Read the upstream agent outputs and evaluate them.\\
You MUST return ONLY this JSON structure --- no other text:\\
\{"issues": ["issue1", "issue2"], "corrections": \{"field1": "corrected\_value"\}, "quality\_score": "<float 0.0-1.0>", "revised\_recommendation": "improved version of upstream recommendation"\}
\end{quote}

\begin{quote}
\ttfamily\footnotesize\raggedright
\textbf{synthesizer:}\\
You are a synthesizer agent. Combine ALL upstream agent outputs into one final answer.\\
You MUST explicitly reference each upstream agent output in your synthesis.\\
Weight conflicting signals: critics $>$ analyzers $>$ planners.\\
Return ONLY valid JSON with the final synthesized output.
\end{quote}

\begin{quote}
\ttfamily\footnotesize\raggedright
\textbf{executor:}\\
You are an executor agent. Apply the plan from upstream agents step by step.\\
You MUST follow the steps from any planner upstream, incorporating corrections from any critic upstream.\\
Return ONLY valid JSON with the execution result.
\end{quote}

\begin{quote}
\ttfamily\footnotesize\raggedright
\textbf{transformer:}\\
You are a transformer agent. Convert upstream outputs into the format the task requires.\\
Read all upstream key\_values and recommendations, then reformat them to match the task output schema exactly.\\
Return ONLY valid JSON matching the task output format.
\end{quote}

\begin{quote}
\ttfamily\footnotesize\raggedright
\textbf{validator:}\\
You are a validator agent. Check the upstream agent output against the task constraints.\\
You MUST return ONLY this JSON structure --- no other text:\\
\{"valid": "true|false", "violations": ["violation1"], "corrected\_output": "<corrected final answer if invalid, else original>"\}
\end{quote}

Four roles --- \texttt{generator}, \texttt{synthesizer}, \texttt{executor}, and \texttt{transformer} --- are deliberately left without a fixed output schema in their initial prompt, since their required output shape is task-dependent (e.g.\ a numeric adoption rate for \textit{mask\_adoption} versus a star rating and review string for \textit{agent\_society}); for these roles the task-specific output format is instead communicated at run time through \texttt{task\_context}, which \texttt{WorkflowTopology.execute} prepends to every agent's prompt regardless of role. Every agent, irrespective of role, additionally receives a final instruction appended by \texttt{GeneralAgent.run} at the end of its prompt: ``Respond with ONLY a valid JSON object. No markdown, no explanation.'' None of these eight prompts are hand-tuned per task; the only task-specific signal an agent receives is \texttt{task\_context} and the upstream outputs passed to it at run time, so any task-specific behavior visible in our results emerges either from this shared context or from the textual-gradient optimization process described above, not from prompt engineering per task.

\section{SOCIA--SWORD Ablation Study}
\label{app:sociasword}
The released implementation is designed to execute and verify generated simulation code inside an isolated Docker container. Because Docker was unavailable in our evaluation environment, the implementation's documented fallback was to ask the underlying LLM to \emph{approximate} the result that code execution would have produced. We do not consider this fallback suitable for evaluation against held-out ground truth, because an LLM-generated execution estimate does not provide a reliable basis for assessing the actual behavior of the generated program.

We therefore modified two call sites in the released implementation: the simulation-execution step and a code-verification smoke test. Both require Docker and provide no independent non-Docker execution path. When Docker was unavailable, we instead executed the generated code as a native Python subprocess. This modification differs from the authors' documented fallback and should therefore be understood as a change to the evaluation protocol rather than a correction to the released implementation. All results reported in this ablation were obtained by genuinely executing the generated programs and evaluating their outputs against held-out data, but under native, unsandboxed execution rather than the original Docker-based execution and verification environment.

Unlike SWORD, SOCIA-$\nabla$ regenerates its implementation from scratch on each invocation and additionally requires the generated code to pass an automated verification stage before execution. Across the evaluation attempts, runs failed either during verification or during execution, including runtime failures in generated data-preprocessing code such as attempts to process non-numeric fields. One additional run was excluded because the framework's evaluator flagged its reported score as ungrounded in the held-out data.

We therefore report the first run that both completed execution successfully and produced a result that was not flagged as unverified. Only 1 of 8 attempts satisfied both criteria. Consequently, these results should be interpreted as single-run point estimates ($n=1$), rather than replicated statistics. This contrasts with the $n=10$ paired trials used for SWORD and limits the strength of conclusions that can be drawn from the SOCIA-$\nabla$ comparison.
\section{Code Repository of SWORD}
\url{https://anonymous.4open.science/r/SWORD-8686/README.md}

\end{document}